# Teorie lunari, effemeridi e sistemi di coordinate

# (Lunar theories, ephemerides and coordinate systems)

**Laura Giannuzzo**

E-Geos (ASI / Telespazio), laura.giannuzzo@external.e-geos.it

This work deals with the comparison of different parameters (French/IMCCE and US/JPL ephemerides) used to calculate the extension of the umbral shadow and the location of the centre line in the total solar eclipse that took place on March, 29th 2006. We needed to know the exact points where the predicted duration of the solar eclipse was 0 seconds. This phenomenon allows, observed on this particular position, the measurement of the solar diameter. The work is composed by three parts. In the first one is described the evolution of the lunar theory from Hipparchus to the current theory (Bureau de Longitudes/IMCCE, Chapront). The second part is dedicated to geodesy (WGS84 datum)and there are examples of longitude's calculations. In the third one there is the analysis of the parameters to understand the reasons about the different extensions of the umbral shadow and the different positions of the centre line in the solar eclipse calculated respectively with French and American ephemerides. The use of the center of the figure of the Moon with respect to the center of mass produces the differences in the ephemerides. This work prepared the observational mission in Egypt to observe the total solar eclipse from the shadow's edges in Zawayet al Mahtallah, near Sidi Barrani, whose results were after published in Solar Physics 259 189 (2009).

**UNIVERSITÀ DEGLI STUDI DI ROMA "LA SAPIENZA"**

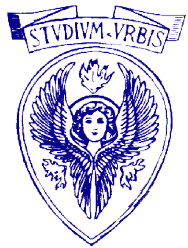

**Facoltà di Lettere e Filosofia**

*Dipartimento di Geografia Umana*

**TESI DI LAUREA DI PRIMO LIVELLO**

in

**GEOGRAFIA**

# TEORIE LUNARI, EFFEMERIDI E SISTEMI DI COORDINATE

Laureanda                                   Relatore

Laura Giannuzzo                Prof. Costantino Sigismondi

Anno Accademico 2004-2005

Tesi discussa il 22/02/2006

# INDICE

Primo Capitolo: Teorie Lunari

## 1.0 INTRODUZIONE

Come definire cos'è una teoria lunare? "Si chiamano solitamente Teorie quei sistemi concettuali che sono l'elaborazione razionale dei risultati delle esperienze". [I]

Quindi le Teorie lunari sono dei modelli analitici o numerici che meglio rappresentano i dati osservati e si presentano poi formule o tabelle a fini di previsione.

## 1.1 ESEMPIO DI MISURAZIONE DEL MESE SINODICO

Il mese sinodico è il tempo impiegato dalla Luna per ritornare allo stesso punto rispetto al Sole. Un esempio di misura è stato fornito da Francesco Bianchini (1662-1729) realizzatore della Meridiana di Santa Maria degli Angeli (1702), in Roma, su richiesta del papa Innocenzo XII, ma realizzata sotto Clemente XI.

Bianchini, infatti, utilizzò la meridiana anche per osservare la Luna e misurare così la lunghezza del mese sinodico. Confrontò le sue osservazioni con osservazioni antiche, in particolare con quelle effettuate da Tolomeo.

Con un telescopio posto in asse con la meridiana, Bianchini per osservare la Luna apriva la finestrella di 60cmx40cm, in cui era alloggiato il foro stenopeico solare, posta nel vertice della meridiana e vi collocava un particolare traguardo che individuasse il meridiano. L'immagine della Luna veniva proiettata a terra sulla linea meridiana, attraverso una diottra[II] ed il tempo veniva misurato con un orologio pendolo.

---

[I] Grande Dizionario Enciclopedico UTET, 1969 III edizione.
[II] Traguardo tubolare senza lenti munito di mirino a croce

Si constata che per tutti il calcolo della lunazione è di 29 giorni 12 h 44m 35sec più una restante differenza di terzi[III]. Nella tabella seguente sono riportati i valori di durata del mese sinodico calcolato dai seguenti autori:

Tabella 0  Valori del mese sinodico misurato da diversi autori[IV]

|            | Giorni | Ore | Minuti | Secondi | Terzi | Quarti |
|------------|--------|-----|--------|---------|-------|--------|
| Ipparco    | 29     | 12  | 44     | 3       | 15    | 44     |
| Tolomeo    | 29     | 12  | 44     | 3       | 20    |        |
| Bullialdus | 29     | 12  | 44     | 3       | 9     | 37     |
| Copernico  | 29     | 12  | 44     | 3       | 12    |        |
| Cal. Greg. | 29     | 12  | 44     | 3       | 10    | 41     |
| Tycho      | 29     | 12  | 44     | 3       | 8     | 30     |
| De Chales  | 29     | 12  | 44     | 3       | 10    | 9      |
| Bianchini  | 29     | 12  | 44     | 3       | 10    | 50     |

L'osservazione di Tolomeo, a cui si riferisce Bianchini per il suo calcolo avvenne nell'anno 125 (secondo il calendario Giuliano) il 5 aprile alle ore 7 dopo il mezzogiorno di Roma (quindi alle 19). Theodor von Oppolzer sostiene che alle 9 pm di Alessandria (18:57 TU) del 5/4/125 l'eclissi di Luna fu osservata da Teone d'Alessandria e poi trasmessa a Tolomeo [V].;von Oppolzer conferma la bontá del dato tolemaico usato da Bianchini per il suo calcolo del mese sinodico.

Quella di Bianchini invece avvenne nel 1699 il 5 marzo secondo il calendario Giuliano (il 15 secondo quello Gregoriano) alle ore 8.06 dopo il mezzogiorno di Roma (20.06).

Pertanto trascorrono 1573 anni Giuliani 334 giorni 1 ora e 6 minuti. La somma contiene 574872 giorni interi più un'ora: ossia 13796929 ore intere che diventano 827815746 minuti, che diventano 49688944760 minuti secondi che moltiplicati per 60 danno 2980136685600 minuti terzi. Questa somma di minuti che si calcola in 1573 anni 334 giorni un'ora e 6 minuti dall'eclisse osservata da Tolomeo e quella osservata da Bianchini deve essere divisa per la durata di un mese sinodico che è definita da tutti gli astronomi di 29d.12h.44'.3'' e circa 11'''. Deve pertanto essere risolta la durata del mese sinodico per i minuti terzi.

---

[III] Sono i sessantesimi di secondo, così come i quarti sono i sessantesi di secondo di terzi secondo l'uso antico.
[IV] Bianchini Francesco De Nummo et Gnomone Clementino, Roma 1703
[V] Oppolzer Theodor von Canon der Finsternisse AK dWiss, Wien, Denkschriften, 1887

Si stimano 29 giorni (che contengono 24 ore l'uno) quindi si hanno 696 ore + 12 ore si hanno le ore complessive contenute nel mese sinodico ossia 708. Moltiplicando le ore per 60 si hanno 42480 minuti ai quali se ne aggiungono 44. Quindi i 42524 minuti complessivi vengono moltiplicati per 60 e danno 2551440 secondi ai quali se ne aggiungono 3. La somma totale è di 2551443 secondi di una lunazione. Volendo ottenere i terzi bisogna di nuovo moltiplicare per 60 si ottiene 153086580 a cui si aggiungono 11 terzi assunti nel mese dall'inizio. L'intera lunazione è di 153086591 minuti terzi. Dall'osservazione di Tolomeo a quella di Bianchini sono trascorse 19467 lunazioni, ognuna delle quali è formata da 153086591 minuti terzi con all'incirca 18/19 cioè 29 giorni 12 ore 44 minuti 3'' e 11''' 57 e 1/3 [VI].

## 1.2 LA LUNA MEDIA MOTO CIRCOLARE UNIFORME

Si inizia la trattazione delle teorie lunari partendo dal moto più semplice: quello circolare uniforme.

Si consideri la Luna un corpo che compie un'orbita perfettamente circolare intorno alla Terra. Essa compie un giro completo (360°) intorno alla Terra in 27.32 giorni.

Ciò significa che se si osservasse la posizione della Luna tutti i giorni si noterebbe che essa si sposta nel cielo di un angolo pari 13.18° (approssimazione alla seconda cifra decimale) ogni giorno. Questo però non avviene neanche considerando il moto della Luna uniforme. Perché?

Rispetto all'eclittica l'orbita della Luna risulta una sinusoide che interseca l'eclittica nei nodi. Guardando la proiezione del tratto di orbita giornaliera della Luna sull'equatore celeste, si nota che ai tropici la Luna é più veloce del moto medio, mentre in corrispondenza dei nodi é più lenta.

Ciò accade perché il piano dell'orbita lunare e inclinato rispetto a quello dell'equatore celeste, e quindi anche considerando il moto lunare come circolare uniforme si hanno lo stesso delle discrepanze rispetto al moto medio.

Vedasi figura 1 alla pagina seguente.

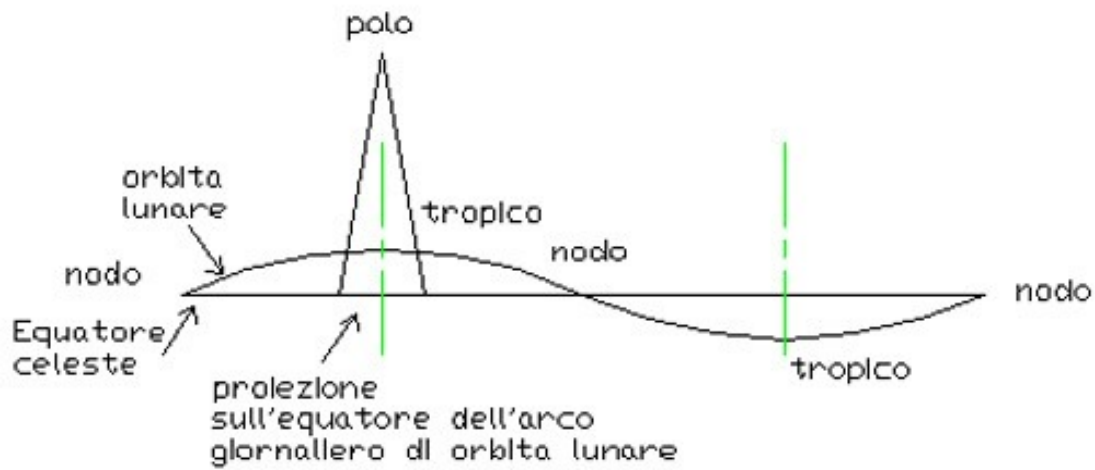

Figura 1. Orbita lunare sull'equatore celeste: Il centro della proiezione è il polo della sfera celeste. Lungo l'eclittica si nota che la proiezione dello spazio percorso giornalmente dalla Luna varia in base alla posizione della Luna lungo la sua orbita. Infatti in prossimità dei nodi la Luna percorre meno spazio sull'equatore celeste che sulla sua orbita, tropici avviene l'inverso. In altre parole la longitudine geocentrica della Luna media aumenta meno ai nodi e più ai tropici. Tutto ciò si traduce in irregolaritá periodiche sugli istanti di transito al meridiano calcolati lì dalle coordinate sull'equatore celeste.

## 1.3 LA LUNA DI IPPARCO

La descrizione del moto successivo riguarda la teoria lunare dell'astronomo ellenico Ipparco, il quale nacque a Nicea di Bitinia ed il periodo della sua attività astronomica va dal 147 a.C. al 127 a.C.

Ipparco nella sua teoria lunare affermò che la Luna ruotava attorno a un cerchio chiamato epiciclo.

L'epiciclo è un'orbita circolare di piccole dimensioni sulla cui circonferenza è collocato o un pianeta o un satellite (figura 2). Esso ruota uniformemente attorno ad un punto T di un'altra circonferenza di raggio maggiore chiamata deferente che a sua volta ruota attorno ad un punto sul quale è collocata la Terra. L'epiciclo e il deferente ruotano entrambi verso Est (moto diretto).

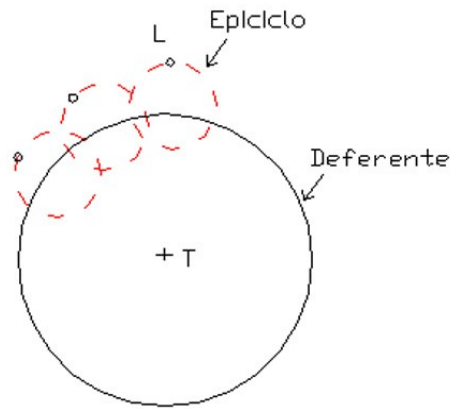

Figura 2 Descrizione di un modello epiciclo-deferente: Le orbite dell'epiciclo sono quelle tratteggiate e il deferente è il cerchio di centro T ( dove è collocata la Terra ) L è la Luna.

Quando il pianeta, per la rotazione dell'epiciclo, si trova all'esterno del deferente, alla massima distanza dalla Terra, il moto dell'epiciclo si somma a quello del deferente, quando si trova all'interno accade l'inverso e il pianeta inverte la sua direzione si ha cioè un movimento di retrocessione in coincidenza con il periodo di massima vicinanza del satellite alla Terra. Per quanto riguarda il moto della Luna intorno alla Terra non c'è moto retrogrado, ma solo differenze di velocità nel tempo rispetto al moto medio.

Esiste un teorema, formulato da Ipparco, ma sviluppato in base a nozioni geometriche ereditate dal suo predecessore Apollonio, che equipara il modello epiciclo-deferente a quello di eccentrico: moto circolare uniforme attorno ad un centro esterno alla Terra. (figura 3).

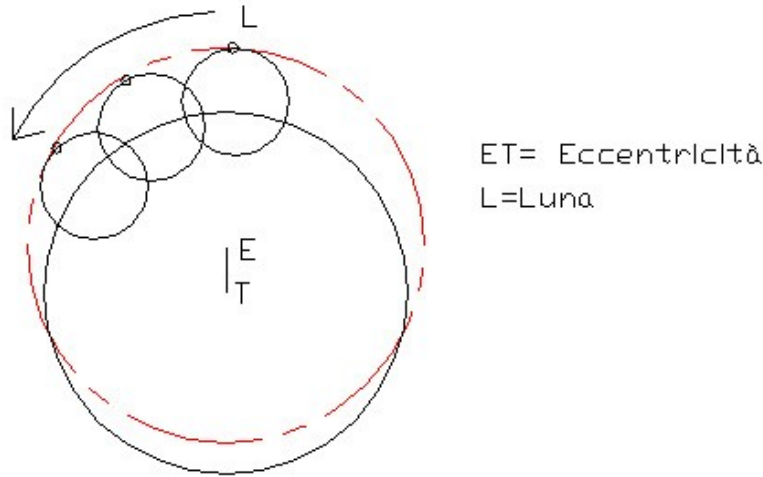

Figura 3 Modello epiciclo deferente equivalente all'eccentrico. ET è il valore dell'eccentricità. Le posizioni della Luna sul suo epiciclo possono essere uniti in una circonferenza di centro E diverso dalla Terra.

Il modello di Ipparco prevedeva un'eccentricità (e) ma per poterlo utilizzare bisognava sapere l'entità di tale eccentricità ed alcune posizioni della Luna dove quest'ultima non risentisse dell'epiciclo (apogeo).

Per poter determinare il valore dell'eccentricità dell'orbita lunare, Ipparco si servì di tre eclissi. La Luna durante le eclissi è posizionata in figura 4 nei tre punti M, N, Q. Gli angoli al centro formati da questi tre punti in C e in T sono noti. Quindi secondo la geometria euclidea si può trovare il rapporto e:R fra l'eccentricità e il raggio dell'eccentrico, che serviva per capire di quanto influisse il valore dell'eccentricità sul moto, ed anche la direzione dell'apogeo relativamente ad una delle tre eclissi. Gli angoli in E sono dati dalla vera posizione della Luna durante le eclissi (ciò non deriva da osservazione diretta di Ipparco ma da archivi Babilonesi). In quel momento la Luna è direttamente opposta al Sole; così egli calcolò la posizione solare in questi tre momenti e per sapere la posizione della Luna ci aggiunse 180°. Gli angoli in C sono ricavati dal moto medio della Luna tra le eclissi, che Ipparco derivò dalle relazioni periodiche assunte in blocco dagli astronomi Babilonesi e che costituiscono un elemento fondamentale per la sua teoria lunare. Tali relazioni sono state attribuite ad Ipparco da Tolomeo ed esse sono:

In 251 mesi la Luna compie 269 ritorni in anomalia in longitudine geocentrica.

In 5,458 mesi la Luna compie 5,923 ritorni in latitudine.

La lunghezza del mese sinodico è 29 giorni, 12 ore, 44', 3", 15''', 44$^{VI}$.

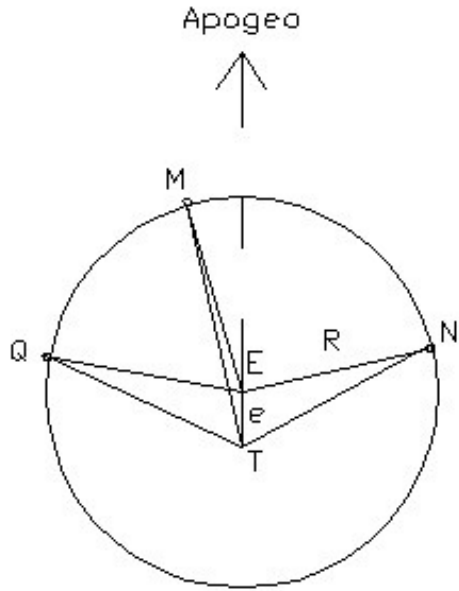

Figura 4 Determinazione di Ipparco dell'eccentricità lunare e della posizione dell'apogeo in base a tre eclissi lunari. A partire dagli angoli formati dalla Luna nelle tre posizioni rispetto al centro E e alla Terra T, si possono calcolare il rapporto e:R e la direzione di ET.[VI]

## 1.4 LA LUNA DI TOLOMEO

Tolomeo Claudio svolse la sua attività ad Alessandria dal 125 d.C. al 175 d.C. circa. Il suo trattato astronomico fu l'Almagesto composto da 13 libri pubblicato intorno al 150 d.C. Il quarto ed il quinto libro dell'Almagesto furono dedicati alla teoria lunare, nel quarto riprende ciò che aveva osservato e calcolato Ipparco, mentre nel quinto propose una radicale modificazione della teoria lunare.

Tolomeo considerava validi i moti medi proposti dal suo predecessore, accettava il metodo di Ipparco per la determinazione dell'eccentricità lunare, allo scopo di determinare di nuovo l'eccentricità o il raggio dell'epiciclo. L'ammontare dell'eccentricità era 5 ¼ : 60, che costituiva un miglioramento rispetto a Ipparco, e nello stesso tempo mostrava che non vi erano variazioni di eccentricità nel corso dei molti secoli trascorsi tra le sue osservazioni e quelle di Ipparco. Un ulteriore vantaggio derivante dalla risoluzione del problema attraverso il confronto tra due terne di eclissi distanziate temporalmente fra loro era che Tolomeo poteva servirsi dei suoi risultati per verificare i moti medi rispetto alla latitudine eclittica e all'anomalia, e riuscì ad apportare una

---

[VI] Walker Christopher L'astronomia prima del telescopio, Dedalo Bari 1997, pag 98-101

piccola correzione alla seconda. Fino ad allora Tolomeo aveva solo adottato e rivisto le procedure ed i risultati di Ipparco, ma come detto prima, nel quinto libro si accorse che il semplice modello di Ipparco funzionava bene solo alle sizigie[VII], però spesso presentava considerevoli differenze rispetto alla posizione lunare osservata in occasione di altre elongazioni fra la Luna ed il Sole, e che tale scarto era massimo in quadratura[VIII], dedusse che l'epiciclo della Luna (di centro C) pur continuando a ruotare uniformemente rispetto alla Terra (O), è trascinato su un circolo il cui centro non è la Terra, ma un punto ulteriore (M), che a sua volta ruota nella direzione opposta lungo un cerchio di centro O. Il moto uniforme di M è il moto medio in elongazione tra il Sole e la Luna (n). Ciò significa che, poiché l'elongazione aumenta a partire dalla sizigia, l'epiciclo è spinto verso la terra dal "gomito" OMC, "connesso" in M e di lunghezza OM+MC pari al raggio deferente nel modello elementare di Ipparco.

Vedasi la figura a pagina seguente.

---

[VII] Sizigie dal greco Sizigie Syn= insieme Zygon=grogo stesso punto delle orbite si intende opposizione al Sole e congiunzione col Sole.
[VIII] Primo ed ultimo quarto.

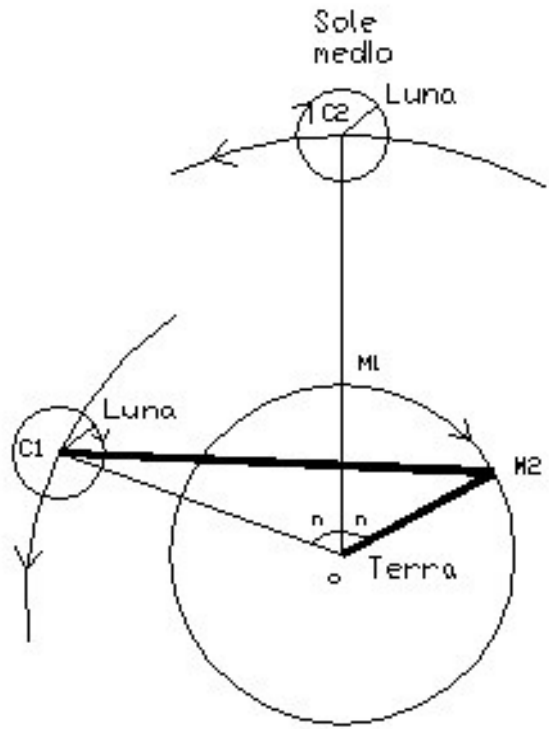

Figura 5 Modello tolemaico perfezionato. L'epiciclo è trasportato su un cerchio il cui centro M, a sua volta si muove su un circolo attorno alla Terra O in senso opposto, seguendo il moto medio in elogazione tra Luna e Sole. L'epiciclo (di centro C) è mostrato in due posizioni. Nella prima, si trova alla massima distanza dalla Terra, mentre nella seconda il "gomito" l'ha spinto verso la minima distanza dalla Terra, e, quindi apparirà molto più grande.

Il modello di Tolomeo rappresenta molto meglio le longitudini, tenendo conto anche di un fenomeno che attualmente è conosciuto come Evezione[IX]: importante disuguagliaza del movimento della Luna con periodo di 32 giorni ed ampiezza di 1°16'. Tuttavia, un grande svantaggio consiste nel fatto che il "gomito" aumenta di molto l'ampiezza dell'intervallo entro cui può variare la distanza della Luna dalla Terra, in modo tale che la sua minima distanza è un po' più della metà di quella massima: ciò dovrebbe manifestarsi in una simile variazione osservabile è molto più ridotta[X]. Secondo la sua teoria quando la Luna si trova in quadratura, e al tempo stesso nel perigeo dell'epiciclo, dovrebbe presentare un diametro apparente grande circa il doppio di quando si trova in congiunzione o in opposizione, se non si vuole ammettere che in tale intervallo di tempo il raggio dell'epiciclo potesse aumentare e diminuire di un fattore pari a 2.[XI]

---

[IX] Lo chiamó cosí Ismael Buillaldus (1605-1694)
[X] Walker Christopher, L'astronomia prima del telescopio, Dedalo, Bari, 1997 pag (106-112)
[XI]. Verdet Jean Pierre, Storia dell'astronomia, Longanesi, Milano, 1995 pag (52-53)

# 1.5 LA LUNA DI NEWTON[XII]

Dopo la scoperta importante dell'Evezione da parte di Tolomeo il nobile danese appassionato di astronomia, a tal punto da dedicarle tutta la vita, Tycho Brahe (1546-1601) scoprì un'altra perturbazione del moto lunare: la Variazione (un'armonica di ampiezza 35minuti d'arco e periodo di circa 15 giorni) essa si annulla alle sizigie ed alle quadrature, mentre è massima agli ottanti[XIII].

Isaac Newton (1642-1727) trattò della sua teoria lunare nei Matematica Principia Naturalis (1687).

Newton basò la sua teoria, come quasi tutta la sua dinamica, su lemmi matematici. Egli applicò i processi di limite della geometria euclidea per ottenere risultati riguardo la tangente e la curvatura di curve che potè usare nella sua dinamica, il suo metodo è simile alla geometria differenziale del XIX secolo.

Newton mostrò che l'incremento dr (infinitesimo) della posizione del vettore r di un punto che si muove su una curva al di sopra di un arco di lunghezza ds è uguale a $1/2Kds^2$ dove K è la curvatura. Se la corrispondente accelerazione lungo il raggio vettore è a e l'incremento dr nel tempo dt è $1/2\alpha dt^2$, l'accelerazione risulta proporzionale alla curvatura (per l'uguaglianza $1/2Kds^2 =1/2\alpha dt^2$ equazione del moto uniformemente accelerato considerando nulla la velocità iniziale).

Newton trattò per lo più di accelerazioni piuttosto che di forze, ed egli espresse la perturbazione dell'orbita lunare rispetto al Sole come una perturbazione dell'accelerazione della Luna verso la Terra.

Se il raggio dell'orbita lunare è "a" e la distanza del Sole dal centro di massa (punto dove si ipotizza che sia concentrata tutta la massa di un corpo o di un sistema di corpi) del sistema Terra-Luna è "R" (figura 6), l'accelerazione media della Luna verso la Terra è $an_L^2$, dove $n_L$ è la velocità angolare media della Luna verso la Terra e l'accelerazione media del centro di massa del sistema Terra-Luna è $Rn_S^2$ dove $n_S$ è la velocità angolare media del Sole attorno alla Terra e alla Luna. La massima accelerazione della Luna rispetto al Sole è $(R+a)n_S^2$, da cui la perturbazione dell'accelerazione

---

[XII] Cook Alan Successes and failure in Newton's lunar theory, Astronomy and Geophysics, Blackwell publishing, 2000, vol. 41, pag.(6.21-6.25).
[XIII] Cesare Barbieri Lezioni di Astronomia Zanichelli, Bologna, 1999

della Luna verso la Terra è proporzionale alla differenza a$n^2_S$ o ($n_S/n_L$)$^2$ volte l'accelerazione della Luna verso la Terra.

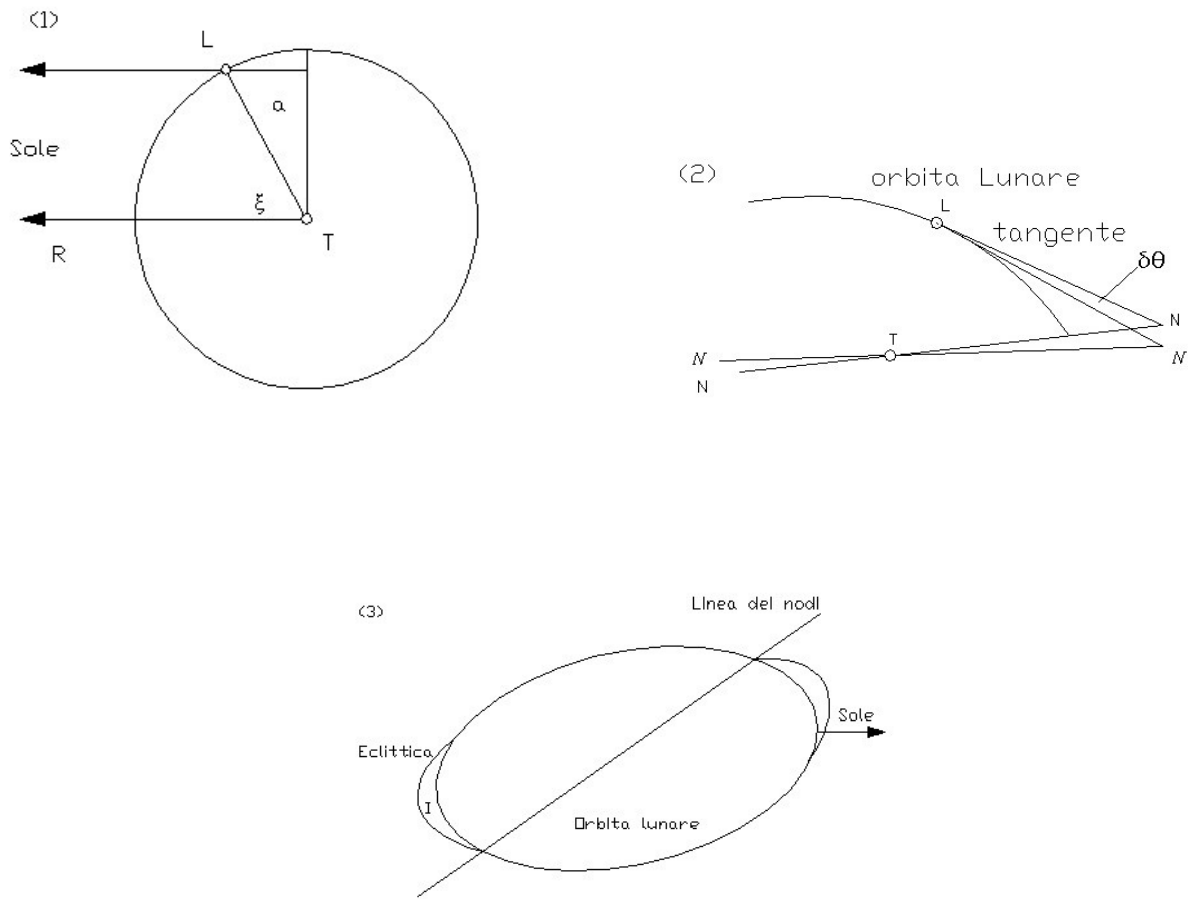

Figura 6 La grandezza della perturbazione dell'accelerazione causata dal Sole(1); La disposizione della tangente all'orbita lunare(2); La dipendenza della forza solare sull'inclinazione(3).(1)-T è la terra, M è la Luna, a è il raggio dell'orbita lunare, R è il raggio dell'orbita terrestre, ξ è l'angolo tra la direzione del Sole e la Luna. L'accelerazione netta della Luna verso il sole è arcosξn2S .(2)-T è la Terra, M è la Luna, NN è la linea dei nodi, NN è la linea dei nodi ruotata, δθ è la rotazione della tangente all'orbita in dt. (3)-I è l'inclinazione del piano dell'orbita lunare sull'eclittica e sull'angolo tra il Sole e la linea dei nodi (N,N).

Il rapporto $n_S/n_L$ è solitamente denominato con m, che è un parametro fondamentale della teoria lunare. Le perturbazioni solari principali sul movimento lunare sono proporzionali a $m^2$ che è pari a circa 1/178. Questo valore può sembrare piccolo ma è sufficiente a creare considerevoli difficoltà nelle soluzioni analitiche dei problemi lunari.

Newton risolse questa rete di attrazione del Sole sulla Luna in componenti lungo il raggio vettore agli angoli retti al raggio vettore nel piano dell'orbita, e perpendicolare al piano dell'orbita.

Alle sizigie (quando la Terra, il Sole e la Luna sono allineati) la prima componente è maggiore e la seconda è trascurabile, mentre in quadratura avviene il contrario.

La prima componente riduce la curvatura dell'orbita ma non ha effetti sulla velocità angolare, o momento angolare, mentre la seconda varia il momento angolare ma non la curvatura. La Variazione è la distorsione della forma dell'orbita lunare per mezzo dell'attrazione del Sole, cosicché l'orbita circolare diventa un'ellisse con la Terra al centro e 2 massimi e 2 minimi di velocità angolare (la variazione dell'orbita) invece di un solo massimo e un solo minimo come una semplice orbita ellittica. Le componenti tangenziale e normale producono una variazione della longitudine della variazione dell'orbita. Tenendo conto che la rotazione del Sole intorno alla Terra, il primo termine della perturbazione è proporzionale a sen$2\xi$, dove $\xi$ è la differenza angolare tra la Luna e il Sole.

Newton trovò la più grande differenza della longitudine nella variazione dell'orbita che gli risultava fosse 33'14'' all'afelio e 37'11'' al perielio. L'attuale valore di sen$2\xi$ è di circa 35', e dipende dalla distanza del Sole dalla Terra. Newton ignorò l'eccentricità dell'orbita lunare nella sua trattazione sulla Variazione (l'orbita lunare differirebbe da un'orbita circolare in misura minore in assenza del Sole) esso infatti esercita sempre la sua attrazione riducendo il valore della forza media esercitata sulla Luna dalla Terra; Newton calcolò questa differenza.

L'orbita della Luna giace in un piano inclinato rispetto al piano dell'eclittica di circa 5°. La forza del Sole sulla Luna, che è parallela all'eclittica, quindi sul piano dell'orbita lunare, tale forza si scompone nelle due componenti normale e tangenziale. I due piani si intersecano lungo la linea dei nodi, che in conseguenza della componente normale della forza solare, ruota sull'eclittica. Allo stesso tempo l'angolo tra i due piani oscilla.

La tangente all'orbita lunare giace sul piano dell'orbita, ed interseca la linea dei nodi. Poiché l'accelerazione solare è parallela all'eclittica essa dà alla Luna una componente della velocità parallela all'eclittica, essa infatti non rimane lungo il piano dell'orbita originario e non interseca più il piano dell'eclittica lungo la linea dei nodi (N,N) ma lungo un'altra (*N,N*).(figura 6).

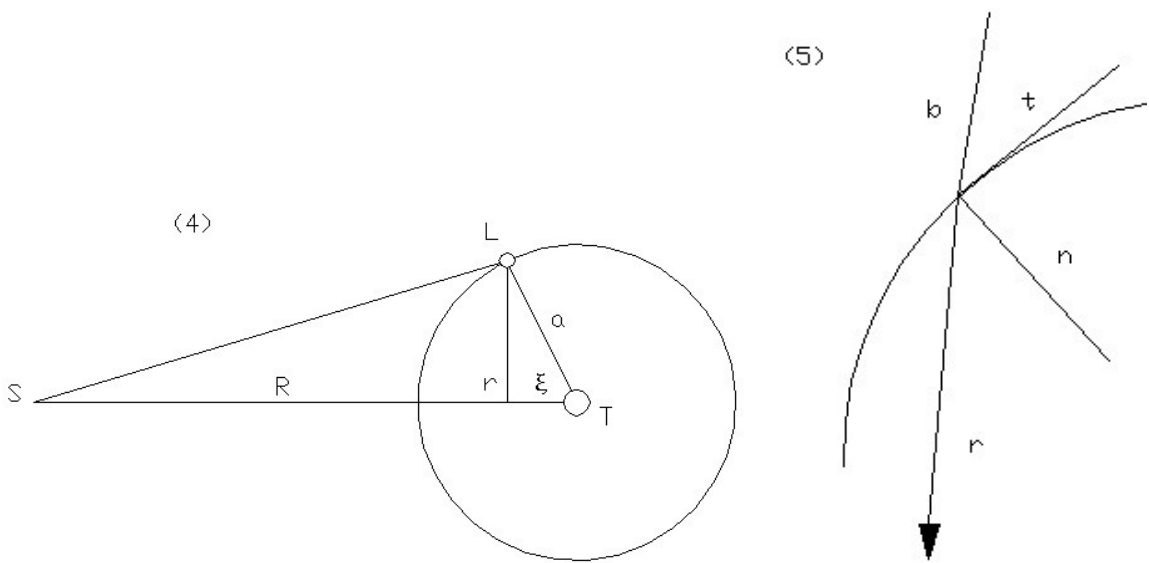

Figura 7 Correzione parallattica (4) e vettori caratterizzanti di una curva (5): (4)-S è il Sole, T è la Terra, L è la Luna, R è la distanza del Sole dalla Terra, $\xi$ è la distanza angolare della Luna dal Sole, a è il raggio centrato dell'orbita della Luna. r è $a\,\text{sen}\,\xi$..(5)-t è il vettore tangente all'orbita, b è il vettore binormale e n è quello normale.

Newton mostrò che la massima velocità oraria della regressione dei nodi era 33''10'''33$^{\text{iv}}$12$^{\text{v}}$, e che il punto medio su un'orbita circolare e la posizione del Sole era circa pari alla metà 16"35'''16$^{\text{iv}}$36$^{\text{v}}$. Newton estese il suo trattato all'orbita ellittica della Luna e trovò che in media il movimento annuale dei nodi era 19°18'1".3, un valore molto vicino a quello calcolato nelle teorie moderne.

L'attrazione del Sole fa ruotare il raggio vettore dell'orbita fuori dal piano originario, cambiando l'inclinazione del piano dell'orbita. Mentre il Sole fa ruotare sempre nello stesso senso la tangente dell'orbita, la rotazione del raggio vettore è verso l'eclittica quando la Luna è nuova ed è assente quando è piena. Così l'inclinazione oscilla attorno al valore medio costante di 5°.

Alla fine della sua sezione sul moto lunare, Newton descrisse alcuni piccoli termini che disse di aver trovato grazie alla teoria gravitazionale.

Ci sono delle variazioni annuali delle longitudini della Luna, dei nodi, e degli apsidi. Ci sono due termini semi-annuali che dipendono: uno dalla distanza angolare del Sole dal nodo lunare e l'altro dalla distanza del Sole dall'apogeo lunare. Infine ci sono due termini con periodi approssimativamente mensili dipendenti dalla distanza angolare tra la Luna ed il Sole. Le ampiezze degli ultimi 4 termini sono di pochi minuti d'arco o meno. L'orbita della Terra intorno al Sole è eccentrica cosicché la distanza del Sole dalla Terra e dalla Luna varia durante l'anno. Così anche il periodo della Luna intorno alla Terra varia con le stagioni, portando ad un'annuale variazione della longitudine della Luna rispetto alle stelle fisse. Newton calcolò quel valore annuale con l'equazione del tempo, la differenza vs, la vera anomalia (longitudine) nella sua orbita intorno al Sole, e Ms, la sua anomalia media uguale a $n_s t$ (dove t è il tempo). La differenza proporzionale all'eccentricità dell'orbita solare è per questo un'indicazione della zona di attrazione solare. Il valore di Newton era 11'50'', più o meno, dipendente dall'eccentricità dell'orbita terrestre intorno al Sole. Il valore moderno è 11'9''.

I termini annuali nei movimenti del nodo lunare e del perigeo sono similmente proporzionali al moto medio terrestre intorno al Sole e al moto del nodo al perigeo. Il moto medio del nodo lunare è circa 19° per anno e Newton stimò che il termine annuale dovesse essere circa 9'24''. Il moto medio del perigeo è circa 40° per anno e Newton rilevò inoltre che la variazione annuale era 19'43''. Questi valori sono vicini alle moderne osservazioni e teorie.

Newton determinò 2 termini semi-annuali di longitudine della Luna, uno con argomento del doppio dell'angolo tra il Sole ed il Perigeo lunare. Quando il Sole è sulla linea dei nodi dell'orbita lunare, la forza che agisce sulla Luna è sul piano dell'orbita e conseguentemente la componente nel piano si riduce, come è la perturbazione della velocità angolare media della Luna. L'ampiezza della variazione della longitudine della Luna è proporzionale a $m^2$ e a $(1-\cos I)$ dove I è l'inclinazione dell'orbita lunare rispetto all'eclittica.

$(1-\cos I)$ è circa $4 \times 10^{-3}$ e l'ampiezza della variazione è circa 45''. Il valore di Newton era 49'' al perielio e 45'' all'afelio; un calcolo più minuzioso dà valore 55''.

La dilatazione dell'orbita lunare a causa dell'attrazione del Sole è maggiore quando Sole, Terra e Luna sono in allineamento. così quando il Sole è ad angolo retto rispetto all'asse maggiore dell'orbita lunare, l'effettivo diametro dell'orbita è il "lato retto" e la distanza media del Sole è la distanza dal Fuoco dell'orbita occupato dalla Terra. Quando il Sole è sull'asse maggiore dell'orbita, l'effettivo diametro è ora il maggior asse e la distanza media del Sole è quella dal centro dell'orbita e non dal Fuoco.

L'ampiezza della variazione in longitudine è proporzionale all'eccentricità lunare e al parametro $m^2$. Poiché la forza perturbatrice solare primaria sulla Luna è il secondo termine di un'armonica (il primo termine scompare per il Teorema del centro di massa) l'argomento del diametro medio è due volte l'angolo tra il Sole ed il Perigeo. I valori di Newton dell'ampiezza dell'effetto erano 3'56'' al perielio e 3'34'' all'afelio. Il valore corrente è 3'32''.

Nella sua teoria Newton considerò che la distanza dal Sole fosse così grande che la direzione dalla Luna era parallela a quella della Terra. Infatti l'angolo tra loro è r/R dove R è la distanza del Sole e r è la perpendicolare che parte dalla Luna e arriva sulla direttrice Terra-Sole.

Se si considera l'orbita della Luna circolare, il rapporto è a/Rsen$\xi$ e a/R è 1/396. La forza del Sole sulla Luna così ha termini addizionali che sono proporzionali a a/R e di argomento $\xi$. Essi generano delle perturbazioni parallattiche nel raggio vettore e nella longitudine della Luna. Il valore di Newton per l'ampiezza del termine in longitudine era 2'20'', mentre il valore attuale è 2'22''.Newton trovò un secondo termine di ampiezza simile 2'25 e argomento uguale alla differenza delle anomalie medie del Sole e della Luna, ma non la differenza delle loro longitudini. Il termine corrisponde a quello moderno di 2'28''.

La direzione del Sole dalla Luna dipende dall'eccentricità dell'orbita della Luna sulla Terra e di quella della Terra rispetto al Sole. La forza del Sole sulla Luna conseguentemente ha termini proporzionali al prodotto dell'eccentricità e con argomenti uguali ai multipli della differenza dei moti medi. Il termine più grande con argomento (Mm-Ms) di longitudine ha un'ampiezza vicina a quella del termine parallattico.

Newton diede entrambi i termini con ampiezze vicine a quelle moderne per il termine parallattico uno con argomento $\xi$, e l'altro con argomento (Mm-Ms), ma egli rimosse il vero termine parallattico nella terza edizione dei Principia.

Nella prima edizione dei Principia Newton fece una trattazione elaborata dell'orbita, che ruota intorno ad un asse inerziale, di un corpo soggetto a forze generali dirette verso un centro comune. Sebbene ciò permetta di ottenere un valore della velocità di rotazione media degli apsidi lunari che era circa uguale a quella del nodo, e metà del valore corretto, Newton non cambiò sostanzialmente le sue edizioni successive.

Newton considerò due orbite della stessa forma geometrica, una fissa e l'altra in rotazione relativa alle stelle fisse, i periodi dei corpi in esse erano gli stessi. Le sue affermazioni implicavano che l'orbita in rotazione e quella fissa erano complanari. Egli considerò solo le forze dirette lungo il raggio vettore della Luna dalla Terra. Sebbene le velocità a punti corrispondenti sulle orbite sono le stesse relative agli assi fissi nelle rispettive orbite, esse sono in differenti direzioni, comportando una differenza di momento angolare. Newton calcolò la forza della perturbazione corrispondente alla rotazione. Il suo terzo passaggio fu un esteso calcolo della velocità di rotazione per ogni generale perturbazione di un'orbita ellittica con una piccola eccentricità. Finalmente applicò il suo risultato all'attrazione del Sole sulla Luna, per trovare che la velocità media di rotazione della linea degli apsidi era la metà di quella osservata.

Newton iniziò a rivedere la sua teoria di un'orbita rotante fino dalla prima edizione dei Principia e continuò a fare così pure nella seconda edizione ma non era soddisfatto dei risultati ottenuti.

Sovrappose al progressivo avanzamento degli apsidi sia l' Oscillazione della linea degli apsidi sia la variazione dell'eccentricità; esse combinate insieme danno una perturbazione chiamata Evezione. Newton adottò un modello cinematico che Halley aveva proposto come un'elaborazione di uno schema di Horrocks di un'ellisse rotante (Halley posizionò il centro di rotazione, il fuoco occupato dalla Terra, su un piccolo epiciclo). Newton sembra non aver tentato una derivazione gravitazionale del modello e lui non ha mai dato un valore di Evezione.

Egli considerò che la linea dei nodi fosse la linea di giunzione tra il centro della Terra e l'intersezione della tangente all'orbita col piano dell'eclittica. Un'alternativa sarebbe usare i vettori dell'orbita come definito dalla geometria differenziale. Il vettore tangente t è la derivata prima del vettore posizione rispetto all'arcolunghezza, l'unità normale n è nella direzione della derivata seconda, e l'unità binormale b è nella direzione della derivata terza rispetto all'arcolunghezza. I tre vettori t,n,b sono rispettivamente perpendicolari e il binormale è perpendicolare al piano istantaneo dell'orbita. Esso definisce l'istantanea inclinazione del piano osculante (dell'orbita) e la sua intersezione con l'eclittica definisce la direzione istantanea della linea dei nodi. La velocità di cambiamento di b è dato dalla componente dell'accelerazione fuori dal piano dell'orbita, da cui la velocità istantanea della rotazione della linea dei nodi può essere calcolata. Le due definizioni della linea dei nodi non sono equivalenti.

Nella definizione di Newton il piano del raggio vettore e la tangente necessariamente contiene il centro di massa Terra-Luna. Il piano definito dai vettori t,n,b non è costretto a passare per quel centro e sulla base dell'attrazione del Sole il generale non farà così. Sulla base dell'osservazione della linea dei nodi può essere la linea che unisce i punti successivi ai quali la Luna passa sull'eclittica. Alternativamente può essere trattata come un parametro da determinare in una trattazione numerica di un'orbita. Ci sono problemi simili nel definire le direzioni apogeo e perigeo.

L'altra questione è la forma delle equazioni dei moti come equazioni per le velocità di cambiamento delle variabili come la longitudine del nodo o del perigeo. Le velocità sono funzioni trigonometriche della differenza di longitudini del Sole e della Luna; se tutti gli altri fattori fossero stati costanti sarebbe stato facile calcolare i moti secolari del nodo o del perigeo. L'equazione del nodo include anche la variabile inclinazione, e l'equazione dell'inclinazione include la variabile nodo. Le due equazioni sono strettamente collegate e dovrebbero essere risolte separatamente perché, per esempio, l'inclinazione è sempre piccola, e il movimento del nodo è quasi quella per inclinazione 0. Similmente la piccola eccentricità dell'orbita lunare ha un effetto minore sulla componente della forza solare perpendicolare al piano dell'orbita; è per queste ragioni che la teoria di Newton era così vicina alle moderne teorie ed osservazioni. La situazione si capovolge per quanto riguarda la rotazione della linea degli apsidi. Nel 1750 Clairaut scoprì una teoria che includeva la soluzione numerica delle equazioni differenziali. Egli inizialmente ottenne gli stessi risultati di Newton e pensò che la legge del quadrato inverso dell'attrazione doveva essere supplementata da un termine cubico inverso, ma poi trovò un errore nella sua analisi che considerò una discrepanza. Clairaut assunse che l'orbita della Luna doveva avere l'equazione polare :

$R = \dfrac{k}{(1+ e\cos c\phi)}$ di un'ellisse riferita ad un asse rotante alla velocità (1-c); c è una costante che determinata dall'analisi, che è equivalente all'assunzione di Newton. Quando egli sostituì quell'espressione in un'equazione del tempo come una funzione di longitudine vera, Clairaut ottenne lo stesso risultato di Newton.

Egli capì che doveva prendere il raggio vettore con migliore approssimazione

$$\dfrac{k}{R} = 1 + e\cos c\phi + \beta \cos \lambda \phi + \gamma \cos(\lambda - c)\phi + \sigma \cos(\lambda + c)\phi$$

λ è un'altra costante da determinare.

Egli trovò un risultato soddisfacente per quanto riguarda la rotazione della linea degli apsidi.

La soluzione della curva rotante non è un'ellisse in uno schema fisso (invece di un singolo termine periodico, il raggio reciproco, riferito ad un asse fisso, ha un numero infinito di incommensurabili componenti periodiche e la curva non è chiusa.

Newton impose una condizione inconsistente con le dinamiche, essa può essere solo soddisfatta con una forza inversa al cubo. Inoltre egli considerò solo la componente radiale della forza di disturbo e ignorò quella tangenziale. Ciò è curioso visto che il suo trattato sulla variazione dell'orbita sembrava inconsistente con queste assunzioni. Lì Newton trattò sia della componente radiale sia di quella tangenziale, mentre i suoi diagrammi mostrano l'aggiustamento del moto del Sole intorno alla Terra e non indicano che lui capì che la variazione orbitale non era una curva chiusa. Evidentemente il moto del perigeo non può essere trovato indipendentemente da un'integrazione completa di un'equazione del moto per un'orbita perturbata, in contrasto con la teoria del nodo che è a primo ordine indipendente dalla forma e dalla vicinanza ad un'orbita circolare. Espressioni per le perturbazioni dell'orbita, dell'eccentricità e della rotazione della linea degli apsidi, può essere derivata geometricamente, ma l'integrazione dovrebbe di nuovo includere equazioni collegate e ciò non è semplice.

## 1.6 LA LUNA MODERNA

Oggi si sa che le componenti del moto lunare sono più di 400 armoniche (Bureau des longitude) Le armoniche che compongono il moto lunare si riescono ad individuare con un processo matematico chiamato FFT (Fast Fourier Transform). Una volta applicato questo algoritmo e quindi individuato le armoniche (sapendo quindi le ampiezze e i periodi) si iniziano a riconoscere le principali. La trasformata di Fourier discreta (spesso abbreviata a DFT, Discrete Fourier Transform) stabilisce una relazione biunivoca tra due n-ple di numeri (generalmente complessi). Queste funzioni hanno la loro principale applicazione nello studio degli andamenti temporali dei fenomeni.

Nel caso del moto lunare si passa da una funzione a dominio nel tempo (cioè dove la variabile dipendente è il tempo) ad un'altra che ha dominio nella frequenza (cioè dove la variabile dipendente è la frequenza).

## 1.7 COME FUNZIONA SU EXCEL

Prendendo i dati di longitudine geocentrica della Luna dal programma Ephmvga relativi ad un periodo di tempo abbastanza lungo (step size di un giorno), in modo tale da poter individuare anche le armoniche che hanno periodo annuale (Teorema di Shannon).

Il Teorema di Shannon o Teorema del campionamento definisce la frequenza minima di campionamento di un segnale deve essere almeno il doppio della frequenza massima dello stesso per evitare distorsioni.

Sapendo che la Luna media compie circa 13.18° al giorno si va a sommare questo valore a tutti i termini di longitudine, in questo modo si ottiene la serie di valori per la Luna media.

Una volta sottratto il moto medio da quello reale si effettua l'algoritmo della FFT per passare al dominio della frequenza e individuare (facendo il modulo quadro delle componenti della FFT) lo spettro di potenza del "segnale" della Luna.

## 1.8 LO SPETTRO DI POTENZA

Lo spettro di potenza di un segnale corrisponde alla percentuale della potenza complessiva del segnale derivante da ciascuna componente spettrale.

Lo spettro di potenza individua le perturbazioni periodiche del moto lunare (armoniche) che possono essere riconosciute dalla loro ampiezza e periodo.

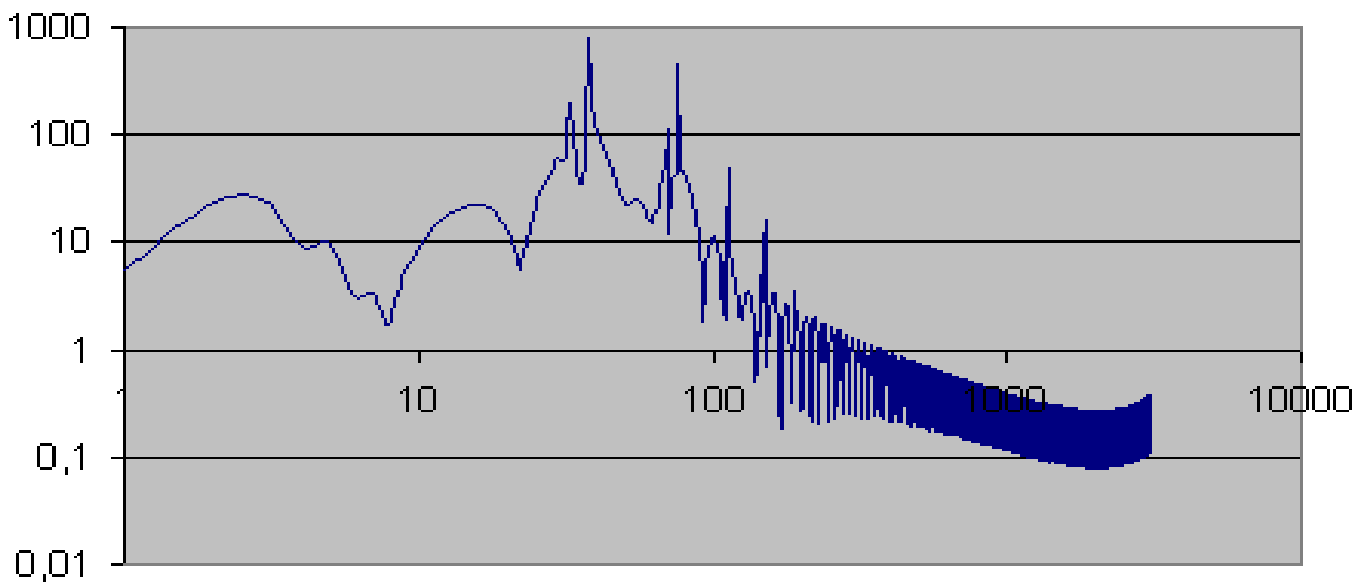

Figura 7 bis Lo spettro di potenza della Luna, in ascissa 1 corrisponde a 1024 giorni, 2 a 512 e cosi' via: si veda la tabella seguente. I picchi appaiono attorno all'anno lunare, al mese e alle "anomalie" più importanti del moto lunare: evezione e variazione.

| | | | | | | | | | | | |
|---|---|---|---|---|---|---|---|---|---|---|---|
| 5,665606 | | 1 | | 72,89633 | 33,03226 | 32 | evezione | | | | |
| 23,20456 | 1024 | 2 | | 201,9329 | 32 | 33 | evezione | | | | |
| 23,55057 | 512 | 3 | | 42,47462 | 31,0303 | 34 | | | | | |
| 8,930346 | 341,3333 | 4 | annuale | 39,33321 | 30,11765 | 35 | | | | | |
| 9,807227 | 256 | 5 | | 34,7627 | 29,25714 | 36 | | | | | |
| 3,124022 | 204,8 | 6 | | 58,22148 | 28,44444 | 37 | | | | | |
| 3,294006 | 170,6667 | 7 | | 762,7968 | 27,67568 | 38 | siderale | | | | |
| 1,709435 | 146,2857 | 8 | | 270,2043 | 26,94737 | 39 | | 25,21497 | 18,61818 | 56 | |
| 5,247665 | 128 | 9 | | 142,9484 | 26,25641 | 40 | | 24,27193 | 18,28571 | 57 | |
| 8,760382 | 113,7778 | 10 | | 112,3914 | 25,6 | 41 | | 22,64449 | 17,96491 | 58 | |
| 12,45542 | 102,4 | 11 | | 91,1955 | 24,97561 | 42 | | 20,49725 | 17,65517 | 59 | |
| 15,81974 | 93,09091 | 12 | | 77,19289 | 24,38095 | 43 | | 18,14129 | 17,35593 | 60 | |
| 18,547 | 85,33333 | 13 | | 65,69369 | 23,81395 | 44 | | 16,10488 | 17,06667 | 61 | |
| 20,59104 | 78,76923 | 14 | | 54,46175 | 23,27273 | 45 | | 15,18136 | 16,78689 | 62 | |
| 21,79226 | 73,14286 | 15 | | 45,19463 | 22,75556 | 46 | | 16,09396 | 16,51613 | 63 | |
| 22,03576 | 68,26667 | 16 | | 36,53957 | 22,26087 | 47 | | 19,02319 | 16,25397 | 64 | |
| 21,26604 | 64 | 17 | | 30,10861 | 21,78723 | 48 | | 23,96219 | 16 | 65 | |
| 19,4606 | 60,23529 | 18 | | 25,3643 | 21,33333 | 49 | | 28,44135 | 15,75385 | 66 | |
| 16,64971 | 56,88889 | 19 | | 22,72037 | 20,89796 | 50 | | 36,08097 | 15,51515 | 67 | |
| 12,95073 | 53,89474 | 20 | | 22,078 | 20,48 | 51 | | 39,19344 | 15,28358 | 68 | |
| 8,720369 | 51,2 | 21 | | 22,74765 | 20,07843 | 52 | | 52,48715 | 15,05882 | 69 | |
| 5,56471 | 48,7619 | 22 | | 23,90433 | 19,69231 | 53 | | 106,3203 | 14,84058 | 70 | variazione |
| 7,792984 | 46,54545 | 23 | | 24,91698 | 19,32075 | 54 | | 12,31151 | 14,62857 | 71 | |
| 13,7589 | 44,52174 | 24 | | 25,41427 | 18,96296 | 55 | | 30,51096 | 14,42254 | 72 | |
| 20,80336 | 42,66667 | 25 | | 25,21497 | 18,61818 | 56 | | 37,86186 | 14,22222 | 73 | |
| 28,25132 | 40,96 | 26 | | | | | | 40,27468 | 14,0274 | 74 | |
| 35,79547 | 39,38462 | 27 | | | | | | 43,57429 | 13,83784 | 75 | |
| 43,2743 | 37,92593 | 28 | | | | | | 447,7661 | 13,65333 | 76 | |
| 50,82837 | 36,57143 | 29 | | | | | | 55,45166 | 13,47368 | 77 | |
| 61,54596 | 35,31034 | 30 | evezione | | | | | 49,47864 | 13,2987 | 78 | |
| 59,26622 | 34,13333 | 31 | evezione | | | | | 47,51876 | 13,12821 | 79 | |

Tabella 1 Spettro di potenza della Luna[XIV]: si identificano le principali componenti armoniche come ad esempio l'Evezione, la Variazione, il mese siderale e la componente annuale attraverso la potenza e periodo delle armoniche.

---

[XIV] Presentzazione PPT, Sigismondi Costantino, La Teoria Lunare 2005

# Capitolo secondo: Determinazione della Longitudine

## 2.0 INTRODUZIONE

Il reticolato geografico consente di determinare la posizione assoluta di un punto sulla superficie terrestre. Tale posizione è individuata da due coordinate paragonabili a quelle cartesiane latitudine e longitudine geografica. La latitudine è la distanza angolare di un punto della superficie terrestre dall'Equatore e può essere Nord o Sud.

La longitudine invece è la distanza angolare di un punto da un determinato meridiano preso come fondamentale[xv]. La longitudine, infatti, non ha una controparte naturale come l'equatore per la latitudine.

Nella storia, infatti, ci sono stati molti meridiani fondamentali, come quello delle Canarie ad esempio. Il meridiano di riferimento attuale è il meridiano che passa per Greenwich.

## 2.1 UN ESEMPIO DI DETERMINAZIONE DELLA LATITUDINE:

Un esempio di determinazione della latitudine l'ha fornito Francesco Bianchini utilizzando lo gnomone meridiano e polare costruito nelle terme di Diocleziano.

Il punto dal quale si vede il transito della polare è quello contrassegnato dal simbolo della stella nella figura 7 della pagina seguente.

Bianchini effettuò due misurazioni dell'altezza della stella polare (una alle 18.00 di sera e un'altra alle 6.00 del mattino). dopodiché corresse i due valori dalla rifrazione atmosferica e ne fece la media. Il valore così determinato della latitudine della Basilica ammontava a 41°54'27'' un valore simile a quello attuale 41°54'11'' (Google Maps).

---

[xv] Bruno Accordi, Elvidio Lupia Palmieri, Maurizio Parotto, Il Globo Terrestre e la sua evoluzione, IV edizione Zanichelli Bologna 1992.

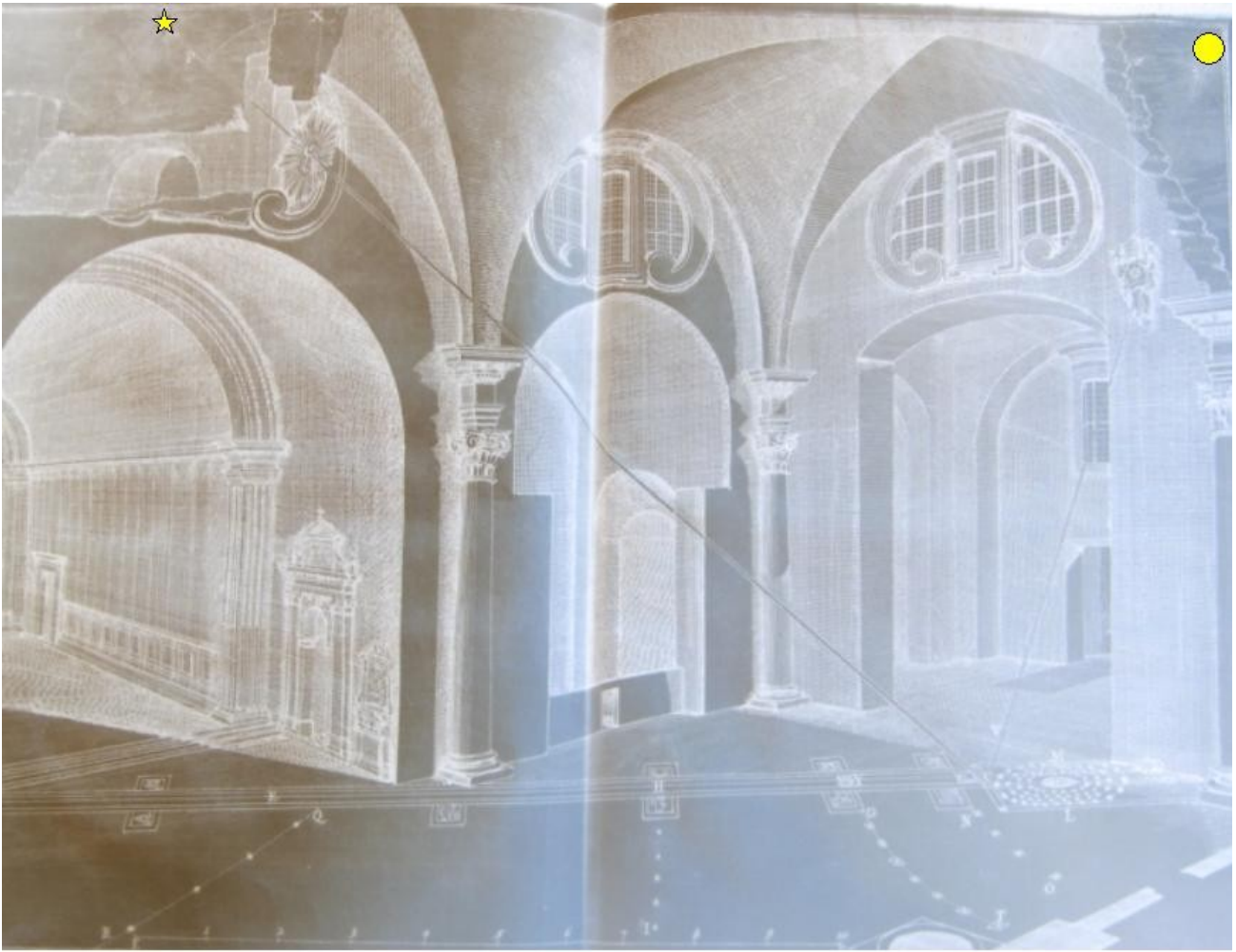

Figura 8, schema di utilizzo della meridiana a Santa Maria Degli Angeli[XVI]: A sinistra è rappresentato il transito della polare, mentre a destra quello del Sole.

## 2.2 UN ESEMPIO DI LONGITUDINE: METODO DELLE DISTANZE LUNARI

Il primo a suggerire questo metodo per la determinazione della longitudine in mare fu Amerigo Vespucci nel 1500.

Successivamente nel 1514 Johann Werner ripropose tale metodo che sarebbe stato conosciuto con il nome di "Metodo delle distanze lunari".

Esso consiste nelle osservazioni dei fenomeni delle occultazioni lunari sfruttando il vantaggio che tali fenomeni hanno di poter essere osservati su tutta la zona della superficie terrestre in cui è visibile la Luna, ma soprattutto utilizza il fatto che la Luna si muove abbastanza rapidamente

---

[XVI] Mario Catamo, Cesare Lucarini, Il Cielo in Basilica, edizione A.R.P.A. Roma 2002.

rispetto alle stelle. Essa, infatti, si muove di una distanza di circa il suo diametro ogni ora e quindi può funzionare come un vero e proprio "orologio celeste". Teoricamente il metodo delle distanze lunari è molto semplice ma richiede che siano soddisfatte tre condizioni:

1-Deve essere nota la posizione delle stelle che vengono interessate dal moto della Luna, con un errore più piccolo di quello che si commette nel calcolo e nella misura della posizione della Luna stessa.

2-Si debbono costruire tabelle (Effemeridi) che prevedano il moto della Luna in cielo, con la stessa precisione.

3-Devono esistere e devono essere facilmente utilizzabili in mare gli strumenti di misura[XVII].

Se si creano tabelle delle distanze della Luna da certe stelle, allora si puo' in principio determinare un tempo assoluto per quel posto di osservazione e determinare così la longitudine paragonando il tempo assoluto col tempo locale. [XVIII]

Johann Tobias Mayer (1723-1762), sovrintendente dell'osservatorio di Gottingen, aveva realizzato la prima serie di tavole lunari sufficientemente precise per poter fungere da base del metodo delle distanze lunari atto a stabilire la longitudine in mare. I marinai dovevano misurare la distanza della Luna da stelle prefissate; tale distanza misurata all'ora locale, opportunamente confrontata con quella riportata sulle tavole, permetteva di ottenere l'ora del meridiano fondamentale cui corrispondeva la medesima distanza lunare. La differenza delle due ore forniva la longitudine cercata. L'Ammiragliato britannico concesse, nel 1765, alla vedova di Mayer parte del premio che doveva essere assegnato a colui che avrebbe risolto il problema della determinazione della longitudine in mare, riconoscendo così anche all'astronomo tedesco, oltre che a Harrison l'inventore del cronometro da marina che manteneva a bordo per mesi l'ora del meridiano di Greenwich, un inizio nella realizzazione di un metodo pratico e affidabile per la determinazione della longitudine in mare. [XIX]

---

[XVII] http://divulgazione.INFM.it/teatro/distanze.htm
[XVIII] http://www.geocities.com/Heartland/Plains/4142/longitude2.html
[XIX] http://www.bo.astro.it/dip/Museum/Old_Museo/italiano/Cronometri.html

## 2.3 ESEMPIO DI CALCOLO DELLE COORDINATE COL METODO DELLE DISTANZE LUNARI UTILIZZANDO IL PROGRAMMA "EPHEMVGA".

Si immagini di essere su una nave, e non sapendo le coordinate in cui ci si trova, si procede a determinarle con il metodo delle distanze lunari.

Si osservi la distanza tra la Luna ed il Sole ad esempio e si misuri questa distanza con uno strumento chiamato sestante.

Il sestante è composto da una scala scorrevole che abbraccia 1/6 di cerchio, da cui il nome, con attaccato uno specchietto orientabile, che permette di dividere in due la visuale: muovendo la scala, si porta il Sole e la Luna a coincidere e contemporaneamente si legge l'angolo tra i due.

Sul programma ephemvga c'é un'opzione (separation) andando nel menu, che dispone i dati in modo tale da leggere le distanze angolari tra i vari corpi celesti date le coordinate terresti, la data e l'orario di osservazione.

Quindi facendo l'operazione inversa, cioé avendo misurato la distanza angolare tra Sole e Luna si modificano le coordinate nel programma in modo tale da far coicidere il dato misurato con quello del programma.

## 2.4 L'EFFETTO DELLA PARALLASSE

La parallasse è l'angolo tra le direzioni sotto cui un oggetto è visto da due diverse posizioni. Nell'esempio rappresentato nella figura 9 si hanno due osservatori: uno posto sul meridiano di fondamentale ove si suppone che stia transitando la Luna, e l'altro posto sul meridiano che passa sull'Etna. A quest'ultimo osservatore la Luna risulterà spostata dell'angolo alfa per effetto della parallasse.

Nella pagina seguente si ha uno schema (figura 9) che visualizza quanto descritto finora.

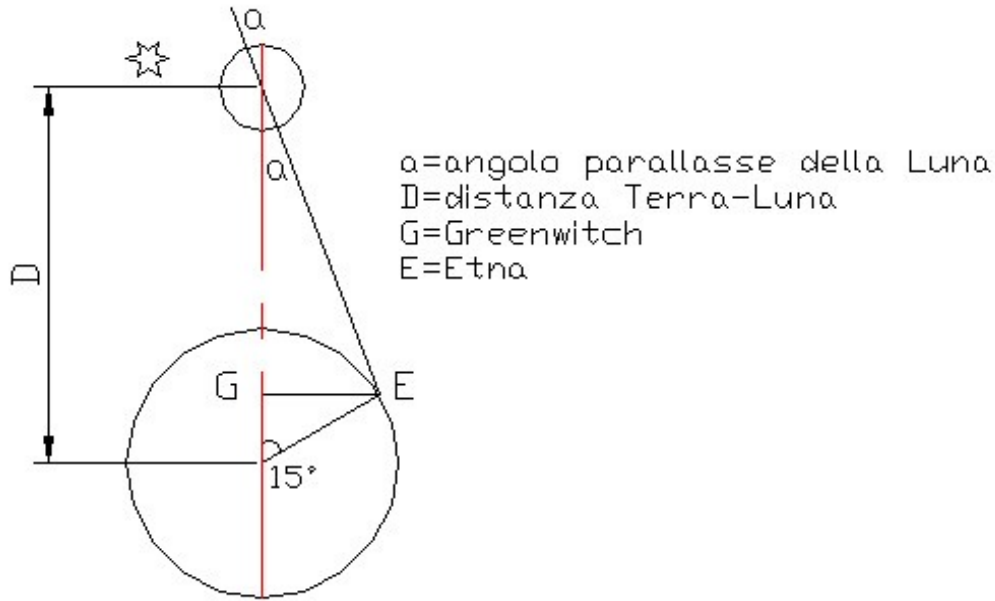

Figura 9 Misurazione dell'angolo di parallasse per una longitudine di 15° Est da Greenwich: Quando l'osservatore posto sul meridiano di Greenwich (G) inizia a vedere l'inizio dell'occultazione lunare, l'osservatore posto sull'Etna (E) riesce ancora a vedere la stella intera perché dal suo punto di vista la luna non ha ancora raggiunto la stella, ciò avverrà dopo 68,31 s.

Per calcolare tale angolo conoscendo il raggio della Terra e la longitudine (15° Est da Greenwich) il segmento perpendicolare al meridiano fondamentale che parte dall'Etna (GE) è dato dalla relazione Rsen(15°) dove R è appunto il raggio terrestre. Una volta determinato GE e sapendo la distanza media della Luna dalla Terra si può calcolare la tg(a) e quindi a che è l'angolo di parallasse:

$Tg(a) = \dfrac{R sen(a)}{D - R\cos(a)}$ da cui facendo l'arctg risultano 0,25° che la Luna media percorre in 68,31 sec. Quindi l'occultazione dall'Etna verrà vista in ritardo di circa 68 secondi rispetto a Greenwich perché la Luna deve compiere ¼ di grado in più.

Se si ripetono queste operazioni per tutti i gradi di longitudine nei quali la Luna è visibile si ottiene una tabella di ritardo dell'occultazione lunare.

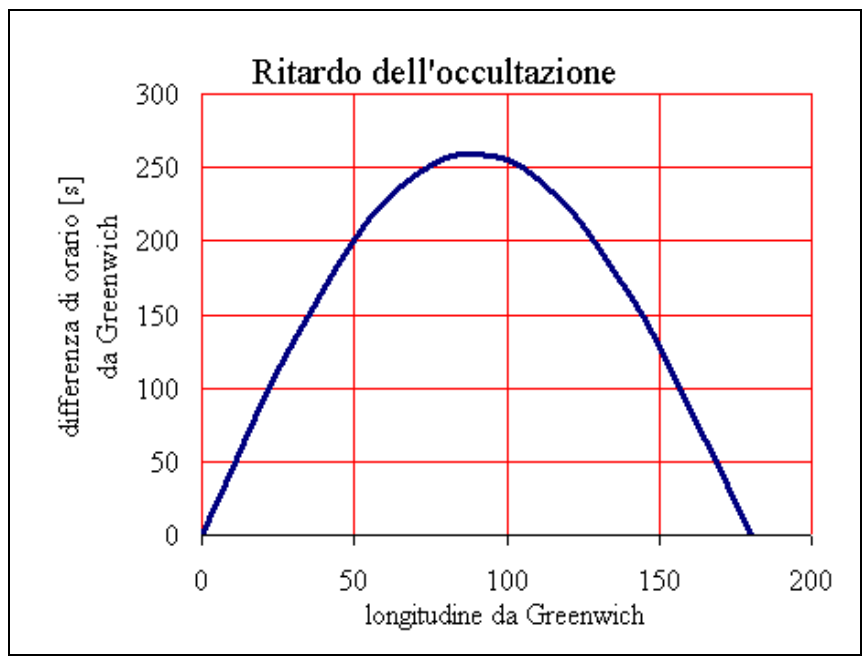

Figura 10: Andamento del ritardo delle occultazioni a causa dell'effetto della parallasse: In questo grafico sull'asse delle ascisse si hanno i valori di longitudine da Greenwich da 0 a 180 e sull'asse delle ordinate la differenza di orario in secondi, causato dall'effetto della parallasse, della vista di un'occultazione di una stella da parte della Luna.

## 2.5 OCCULTAZIONI ALL'US NAVAL OBSERVATORY

Osservando la Luna dalla Terra, se ne può vedere un po' più della metà, grazie ai fenomeni noti come Librazioni.

Come i piatti di una bilancia possono oscillare come un pendolo su e giù attorno alla loro posizione di equilibrio, il moto di librazione della Luna prende il nome da questo tipo di oscillazione a cui rassomiglia. Nella posizione di equilibrio, l'asse maggiore della Luna è puntato verso la Terra, e la librazione fa variare temporaneamente questo puntamento un po' verso nord, sud, est e ovest (librazioni in latitudine e in longitudine). Poiché tutta la Luna segue questo moto, tramite la librazione è possibile osservare un po' di più della sua superficie ( 59,9%).[xx]

---

[xx] http://www-spof.gsfc.nasa.gov/stargaze/Imoon4.htm

All'US Naval Observatory hanno costruito una mappa dei lembi lunari al variare di tutte le librazioni, e quindi, in questo modo hanno ottenuto un valore di massima precisione di longitudine astronomica (0,2'').

Un esempio di librazione è quello nella figura 11.

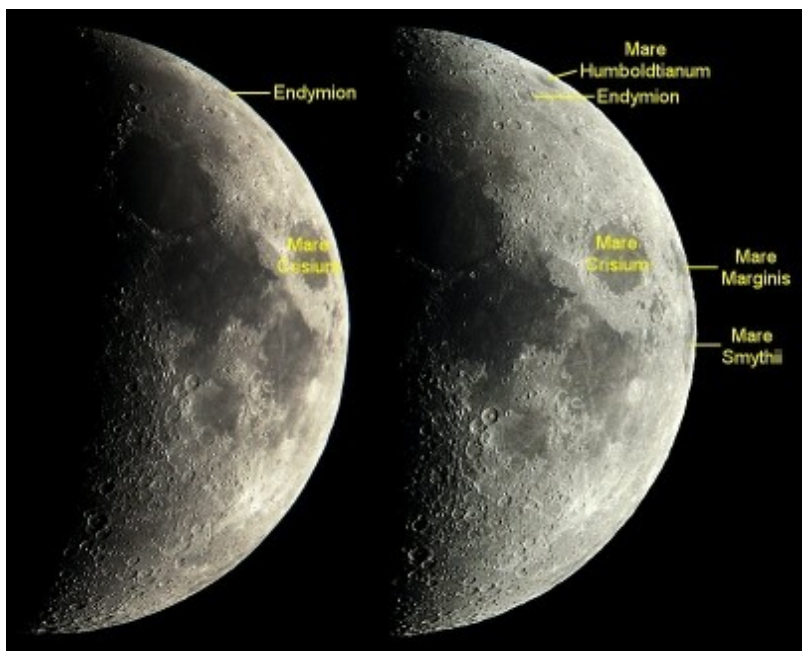

Figura11 Esempio di librazione in longitudine: Nella prima immagine della Luna sono visibili solo il mare Crisium e L'Endymion. Nella seconda, per effetto della librazione sono visibili anche il Mare Smythii, il Mare Marginis e il mare Humboldtianum.

## 2.6 LA VARIAZIONE DELLE COORDINATE A SECONDA DEL SISTEMA DI COORDINATE ADOTTATO

La superficie terrestre è di forma irregolare e non semplice da definire matematicamente occorre quindi "riferire" tutti i punti della superficie terrestre ad un'altra più semplice denominata superficie di riferimento. I punti della superficie terrestre vengono quindi "proiettati" su questa superficie di riferimento che deve avere le seguenti caratteristiche:

1-Deve approssimare bene la superficie terrestre.

2-Deve avere una rappresentazione matematica "semplice".

3-Deve essere possibile stabilire una corrispondenza biunivoca fra i punti della superficie terrestre e quelli della superficie di riferimento.

4-Deve essere possibile istituire una geometria peri calcoli geodetici sulla superficie di riferimento.

La scelta della superficie di riferimento può essere effettuata tra tre tipologie:

lo Sferoide, il Geoide o l'Ellissoide. I punti della superficie terrestre vengono idealmente proiettati sulla superficie di riferimento che si è scelta.

Lo sferoide è una superficie ideale pertanto non indicata per le misurazioni perché è un'approssimazione troppo "grossolana" che non permette di ottenere valori di posizione se non con una precisione di qualche km.

Il Geoide è la migliore approssimazione della superficie terrestre, calcolata a partire dallo studio del campo gravitazionale (è una superficie che unisce tutti i punti aventi la stessa energia potenziale) ed è fisicamente individuabile attraverso il livello medio dei mari in condizioni ideali. Il problema nell'adottare il Geoide come superficie di riferimento sta nel fatto che la sua rappresentazione matematica è particolarmente complessa (è definito in funzione di infiniti parametri).

La superficie di riferimento comunemente usata è l'Ellissoide di rotazione. Esso approssima la superficie terrestre meno correttamente del Geoide e non è fisicamente individuabile ma ha una rappresentazione matematica è semplice. La sua equazione è :

$$\frac{X^2 + Y^2}{a^2} + \frac{Z^2}{c^2} = 1$$

dove a è il semiasse maggiore e c è il semiasse minore dell'ellisse.

I parametri che definiscono l'ellissoide di rotazione sono lo schiacciamento e l'eccentricità che si determinano con le seguenti formule:

$$f = \frac{a - c}{c} \qquad e = \sqrt{\frac{a^2 - c^2}{a^2}}$$ dove f è lo schiacciamento ed e l'eccentricità.

Tabella 2 Alcuni Ellissoidi importanti[XXI]:

| Ellissoide di Bessel | (1830) | a=6377397 m | c=6356079 m | f=1/299.15 |
|---|---|---|---|---|
| Ellissoide di Hayford | (1909) | a=6378388 m | c=6356912 m | f=1/297 |
| Ellissoide WGS84 | (1984) | a=6378137 m | c=6356752 m | f=1/298.257 |

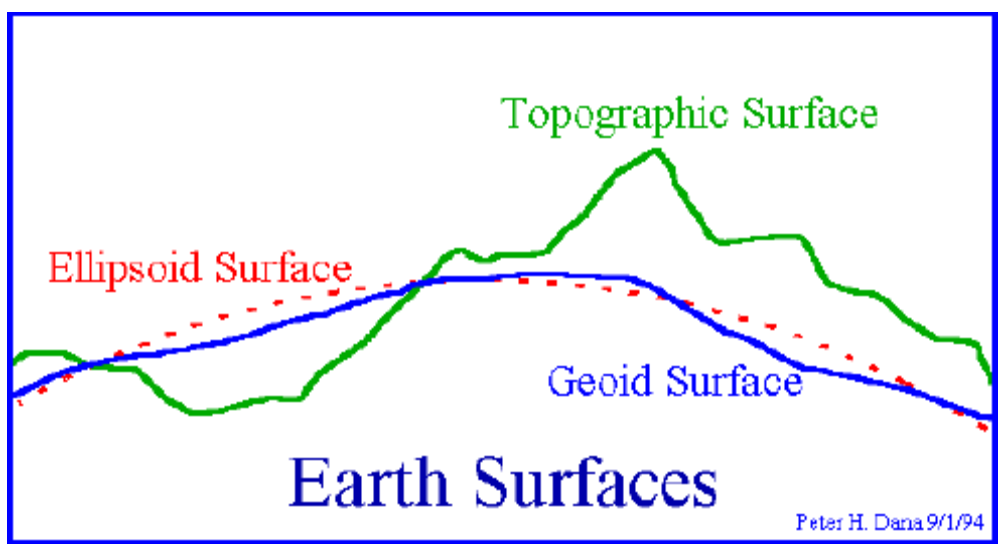

Figura.12 confronto tra la superficie del Geoide, quella dell'Ellissoide e quella topografica[XXII]: La superficie topografica, in prima approssimazione, la si riconduce ad un Geoide (identificato con il livello medio di tutti i mari, cioè come se riempissero tutte le fosse oceaniche e penetri nelle montagne) e in secondo momento ad un determinato Ellissoide che meglio approssima la superficie del Geoide.

La definizione di Datum consiste nella scelta di un ellissoide e di parametri assegnati orientato in modo opportuno rispetto alla Terra. Si distinguono in: datum globali tridimensionali (es.WGS84, del quale si parlerà nel prossimo paragrafo) e datum locali planimetrici (es. Roma40).

Il datum locale ha un campo di validità ristretto, viene riferito ad un ellissoide locale che approssima il geoide solo intorno al punto di emanazione (punto centrale del campo di validità). Essi realizzano l'obbiettivo di poter trasportare, senza eccessivo errore, gli angoli misurati sul terreno direttamente sull'ellissoide.

---

[XXI] http://www.rilevamento.polimi.it/doc/cart_migliaccio/lez%202%20rappresentazione%20della%20sup%20terr
[XXII] http://colorado.edu/geography/graft/notes/datumf.html

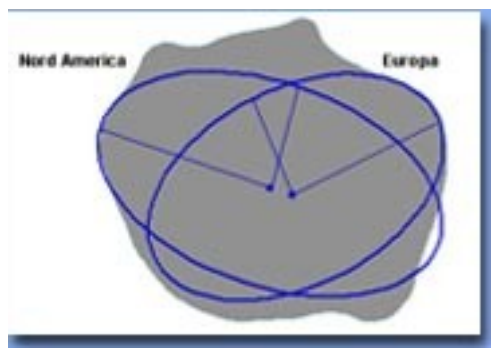

Figura13 Approssimazione del Geoide[XXIII]: Per far aderire il più possibile la superficie ellissiodica con il Geoide vengono utilizzati differenti ellissoidi con diffrenti punti di emanazione, in particolare l'ellissoide di sinistra aderisce meglio al nord America, mentre l'ellissoide di destra aderisce meglio all'Europa.

Nella tabella 3 seguente sono indicati i sistemi di riferimento geodetici (Datum) adottati nel tempo in Italia, mentre nella tabella 4 sono indicate le coordinate della cupola dell'osservatorio di Monte Mario a Roma nei diversi sistemi di riferimento.

Tabella.3 Sistemi di riferimento usati in Italia:

| Denominazione | Ellissoide | Centro di emanazione |
|---|---|---|
| Sistema geodetico nazionale Roma prima del 40 | Bessel | Genova Istituto Idrografico della Marina (ex Osservatorio Astronomico) ed altri |
| Sistema geodetico nazionale (Roma 40) | Internazionale Hayford | Roma Osservatorio Astronomico Monte Mario |
| European Datum 50 (Ed 50) | Internazionale Hayford | Postdam Orientamento Medio Europeo |
| World Geodetic System 84 (WGS 84) | GRS | Coincidenza dei centri ellissoide e geoide (geocentrico) |

---

[XXIII] http://www.analisidifesa.it/numero17/do-geo2.htm

Tabella 4 Coordinate di Monte Mario nei tre sistemi di riferimento

| Datum | Latitudine | Longitudine |
|---|---|---|
| Roma before 40 | 41° 55' 24, 399'' | 0° = 12° 27 ' 06, 840'' da Greenwich |
| Roma 40 | 41° 55' 25, 510'' | 0° = 12° 27 ' 08, 400'' da Greenwich |
| Ed 50 | 41° 55' 31, 487'' | 0° = 12° 27 ' 08, 933'' da Greenwich |

Risulta abbastanza evidente che cambiando ellissoide,e quindi sistema si riferimento, cambiano anche i valori delle coordinate. Come mostrato in figura 13

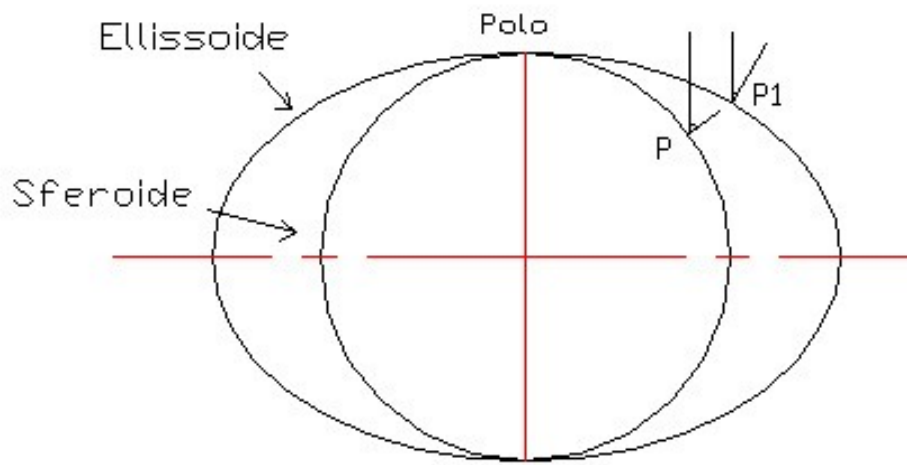

Figura14 Esempio di variazione del valore della latitudine al variare del sistema di riferimento: Il punto P sulla sfera ha un valore di co-latitudine pari a circa 50°. Lo stesso punto proiettato sull'ellissoide ha valore di co-latitudine di circa 30°. Quindi più il sistema di riferimento presenta schiacciamento maggiore e più un punto sulla sua superficie sembra essere a più a Nord rispetto alla latitudine misurata sulla sfera.

## 2.7 IL DATUM GLOBALE (WGS84)

Con l'avvento dei satelliti è sorta la necessità di un unico sistema di riferimento geodetico su scala mondiale. E' stato introdotto quindi il World Geodetic System (WGS84) basato (come scritto in tabella 3) su un ellissoide geocentrico la cui forma e dimensione approssima il geoide nel suo complesso ed è valido per tutto il mondo.

Il datum locale, come detto nel precedente paragrafo, approssima bene il geoide solo nella zona prossima al centro di emanazione.

La differenza tra datum locale e datum globale si può osservare nella figura 14.

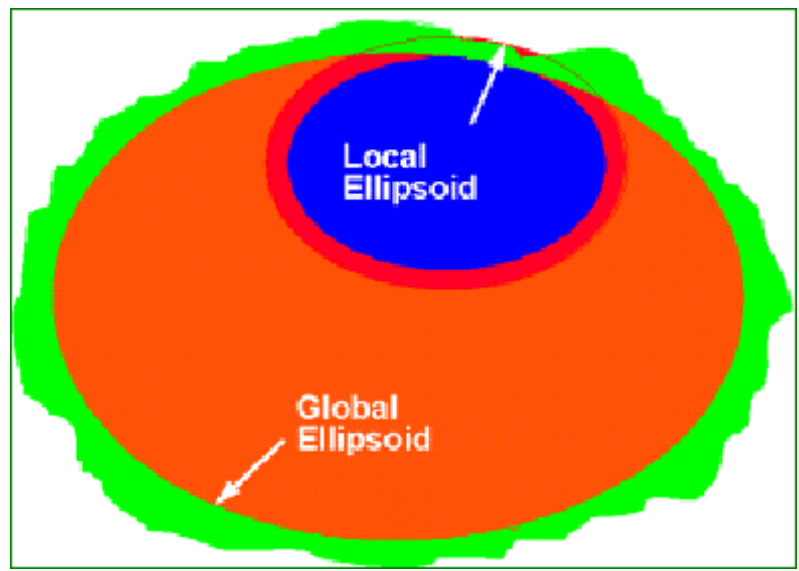

Figura 15 Differenza tra ellissoide locale e globale: Si noti come l'ellissoide locale approssimi bene la superficie del geoide solo nei pressi del punto di emanazione. Il WGS 84 è definito da un insieme di parametri primari e secondari: quelli primari stabiliscono la forma e la massa dell'ellissoide terrestre e la sua velocità angolare; quelli secondari definiscono un modello dettagliato del campo di gravità.[XXIV]

Il sistema di riferimento del WGS84 per le coordinare cartesiane sono:

Z diretto come il Polo medio dell'anno 1984

X diretto come l'intersezione fra il piano equatoriale ed il piano del meridiano di Greenwich,

Y tale da completare la terna destrorsa.

---

[XXIV] http://dipastro.pd.astro.it/planets/barbieri/didattica.html#triennale

# 2.8 I SISTEMI DI COORDINATE:[XXV]

I sistemi di coordinate più utilizzati sono:

1-Coordinate geografiche ellissoidiche

2-Coordinate cartesiane geocentriche o ellissocentriche

3-Coordinate cartesiane locali (euleriane)

4-Coordinate geodetiche polari

5-Coordinate geodetiche ortogonali

A seconda dei tipi, le coordinate possono esprimere una posizione solo planimetrica (coppia di coordinate) o tridimensionale (terna di coordinate che forniscono georeferenziazione plano-altimetriche).

La georeferenziazione é la tecnica di localizzazione territoriale che permette di associare un particolare oggetto ad un punto nello spazio reale.

1- Le coordinate geografiche ellissoidiche sono la latitudine e la longitudine geografiche definite all'inizio del capitolo. Esse da sole definiscono solo la posizione planimetrica, se si associa l'Altezza ellissoidica (distanza del punto considerato dall'ellissoide misurata lungo la normale ellissoidica[XXVI]) si ottiene una terna di coordinate che definiscono la posizione planoaltimetrica del punto.

Si utilizzano per tutte le applicazioni cartografiche.

2- Le coordinate cartesiane geocentriche (Tipologia di coordinate del WGS84) sono una terna di coordinate (X,Y,Z) che hanno:

-origine nel centro di massa della Terra

---

[XXV] http://labtopo.ing.unipg.it/files_sito/compiti/georef.pdf
[XXVI] http://users.libero.it/profLazzarini/geometria _sulla_sfera/geo.htm

-Asse Z diretto lungo l'asse polare medio terrestre -Assi X e Y sul piano dell'equatore (perpendicolare all'asse Z) con X disposto secondo il meridiano fondamentale e Y disposto in modo da formare la terna della mano destra.

Si utilizzano in geodesia satellitare e per fare trasformazioni di datum.

3- Le coordinate cartesiane locali (o euleriane e,n,h figura 1) sono riferite ad una terna euleriana avente:

-Origine in un punto P dell'ellissoide.

- L'asse h diretto secondo la normale ellissoidica passante per P

- Gli assi e ed n sul piano tangente all'ellissoide in P con n diretto secondo la tangente al meridiano verso Nord ed e diretto secondo la tangente al parallelo verso Est.

Si utilizzano in geodesia satellitare e per rilievi locali.

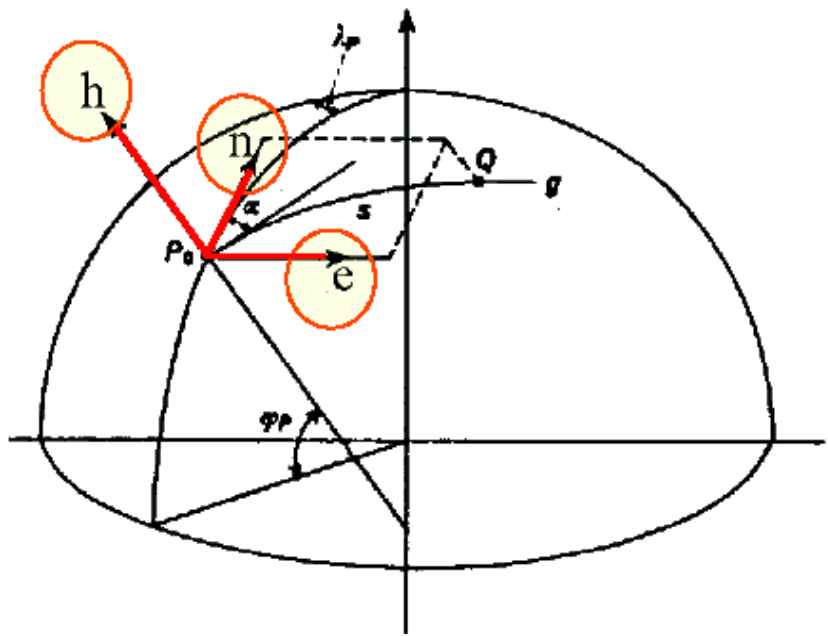

Figura16 Le coordinate cartesiane locali: (e,h,n) del punto P0 corrispondenti alla normale all'ellissoide passante per il punto P0 (h), una tangente al meridiano passante per P0 (n), ed infine la tangente al parallelo passante per P0 (e).

- Le coordinate geodetiche Polari (s,α figura 2) dove s é la distanza polare che è la lunghezza dell'arco di geodetica OP (l'arco di geodetica è la distanza minima che intercorre tra due punti su una sfera, ciò che nel piano corrisponde ad una linea retta) e α è l'azimut in O della geodetica OP

Si utilizzano per calcoli geodetici locali e calcolo di reti

-Le coordinate geodetiche ortogonali (X,Y) dove X è la lunghezza dell'arco di meridiano OQ e Y è la lunghezza dell'arco di geodetica QP ortogonale in Q al meridiano per O.

Sono coordinate di cartografia catastale.

Sia le coordinate (s, α) sia (X,Y) definiscono una posizione planimetrica.(figura 16)

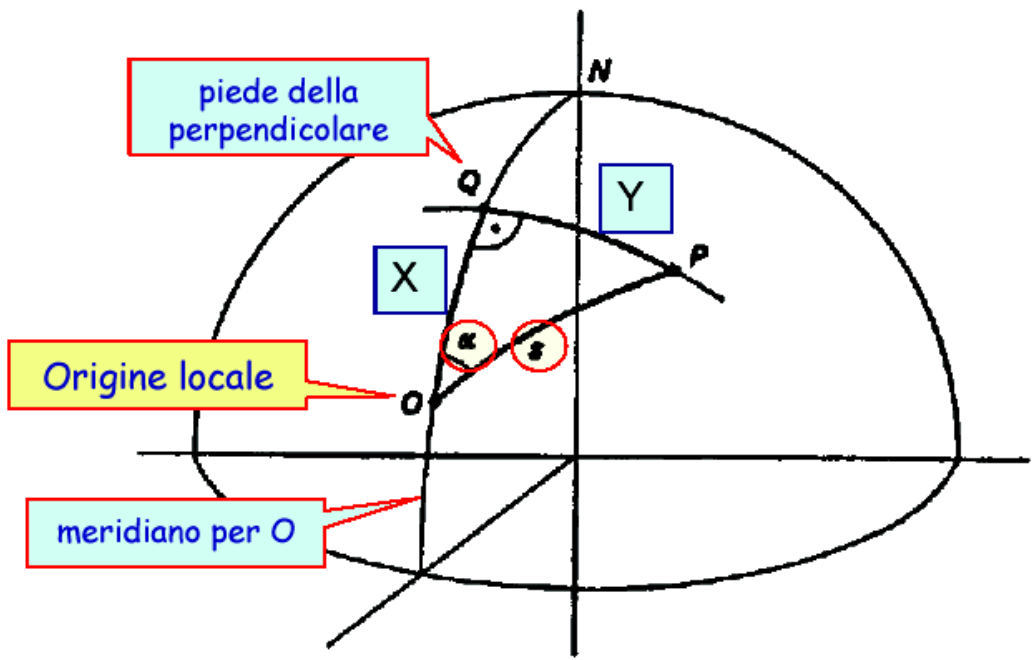

Figura17 Le coordinate geodetiche polari e quelle ortogonali: Sono entrambi coppie di coordinate planimetriche locali (ad esempio il punto O) e si misurano o attraverso i due archi di geodetica (X;Y) oppure attraverso la distanza angolare della geodetica ($\alpha$) ed il valore della sua lunghezza (s). La distanza angolare si misura sempre in senso orario.

# Terzo Capitolo Correzioni delle Effemeridi

3.0 INTRODUZIONE

I dati astronomici variabili nel tempo, come le coordinate del Sole, della Luna e dei pianeti, l'ora del sorgere e del tramontare di questi astri, le fasi lunari, gli istanti di inizio delle varie stagioni in ciascun anno e così via sono riportati in tabelle dette effemeridi, termine di origine greca che significa diario. Le effemeridi vengono pubblicate anno per anno in apposite pubblicazioni dette almanacchi o annuari astronomici.[XXVII]

Oggi, attraverso il GPS (Global Position System), si può tracciare su immagini satellitari (sul programma Google Maps) la posizione sulla Terra di dove sarà visibile un'eclisse di Sole (che è stata precedentemente calcolata nelle previsioni), e da ciò si riescono a migliorare le effemeridi lunari confrontando la previsione con l'eclissi.

Il GPS è un sistema di navigazione d'area globale satellitare. La sua struttura si compone di tre segmenti: spaziale (costellazione di 24 satelliti), di controllo (stazioni a terra), di utilizzazione (ricevitori). Le caratteristiche di tale sistema sono:

-ricezione del segnale 24 h/24 h globale

-ricezione in ogni condizione climatica

-coordinate x,y,z riferite ad un unico sistema di riferimento globale (WGS84).

Ogni satellite conosce la sua posizione e la trasmette nel codice identificativo; ogni ricevitore, calcolando la differenza $\Delta t$ tra la serie di impulsi generata e ricevuta e sapendola velocità della luce, ricava la distanza dal satellite. Ricevendo i dati da tre satelliti per la posizione e da un quarto satellite per sincronizzare gli orologi, si ottiene la posizione a terra.[XXVIII]

3.1 ESEMPIO: ECLISSI DI SOLE 2005 IN SPAGNA

Le eclissi solari, come detto in precedenza; possono contribuire alla correzione delle effemeridi lunari; infatti a seconda delle effemeridi e dei parametri utilizzati per la previsione di tali eclissi si

---

[XXVII] Enciclopedia astronomica Alla scoperta del cielo, Curcio 1982
[XXVIII] CIGA 11° Stage di cartografia operativa.

individuano diverse coordinate geografiche (grazie al programma Google Maps che lavora con il sistema satellitare GPS) del bordo di totalità. Nel caso dell'eclissi del 3 ottobre 2005 in Spagna sono state messe a confronto le previsioni di Espenak con quelle di Herald:

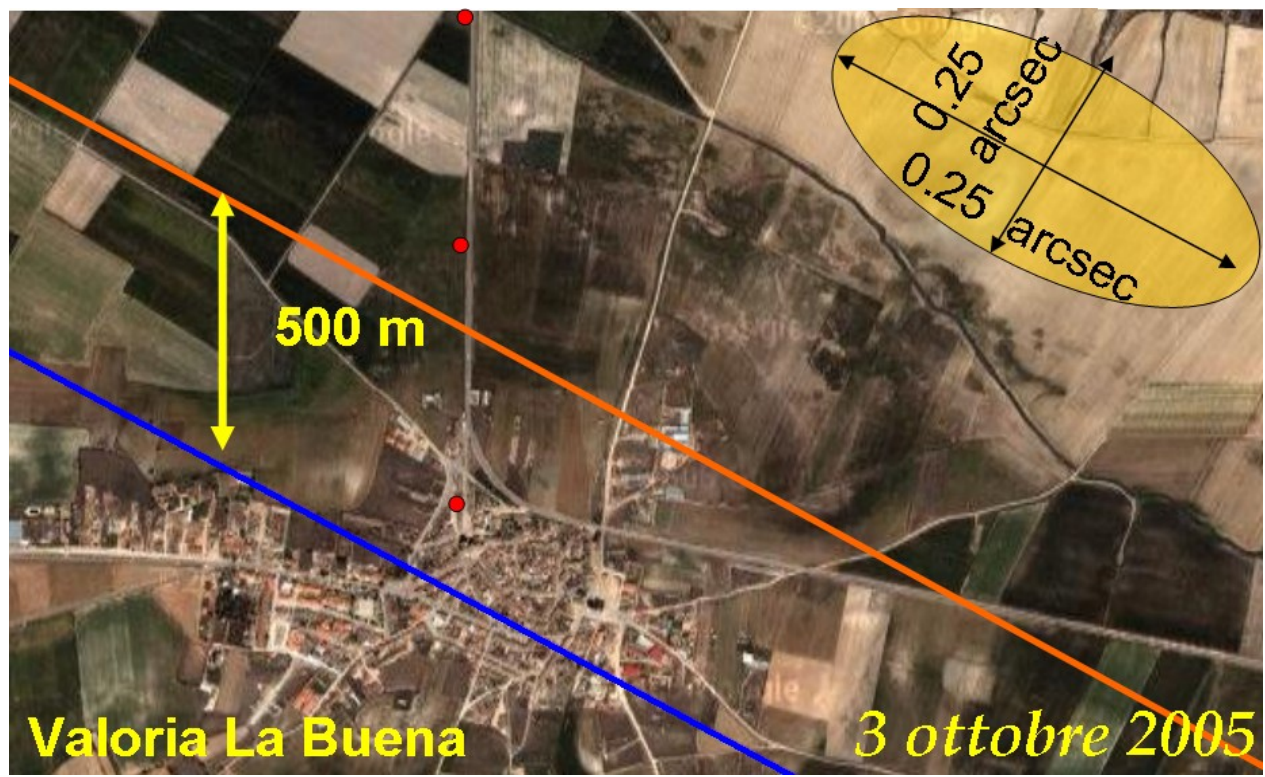

Figura 18 differenze del bordo nord dell'eclisse: Il bordo in blu è quello calcolato da Espenak, quella rossa è stata calcolata da Herald, che ha utilizzato le effemeridi VSOP e per la Luna K=0,272281 (raggio medio lunare misurato in raggi terrestri) e le effemeridi ELP82 di Chapront basate sulle DE200 della NASA. Espenak usa lo stesso k di Herald, le effemeridi solari di Newcomb (1895) e le ILE (1954) per la Luna. L'ellisse rappresenta 0.25 arcsec alla distanza della Luna di 396000 Km calcolata per l'altezza del Sole di 27° al momento di massima eclisse.[XXIX]

Le tavole pubblicate dall'IMCCE (Institut de Méchanique Celeste e de Calcul des Éphémerides) da Rocher basate sulle effemeridi ELP2000 corrette di 0.50 arcsec in longitudine e -0.25 in latitudine (per fornire la posizione del centro ottico che non coincide con quello di massa), mentre per la Terra ed il Sole sono state usate le SLP98 e il k standard (quello consigliato dall'UAI). Nel caso dell'eclissi Spagnola il bordo Nord calcolato dall'IMCCE passa 6.8 Km più a Sud di quello di Espenak, mentre il bordo Sud risulta 3.7 Km più a Sud. Le diverse effemeridi, correzioni per il centro ottico della Luna, per la rifrazione atmosferica, ed anche per il diverso ellissoide di

---

[XXIX] Sigismondi Costantino, Oliva Pietro, Osservare l'eclissi solari dal bordo della fascia della totalità, articolo dell'Astronomia UAI 2006

riferimento per la Terra producono uno spostamento rispetto alle previsioni NASA e WinOccult; inoltre la fascia calcolata dalla NASA è complessivamente più stretta di 3.1 Km dovuti all'uso del k standard.

Il miglioramento delle effemeridi viene effettuato inserendo nel programma WinOccult modifiche del baricentro lunare di posizione in ascensione retta e declinazione in sede di simulazione o analisi dei dati.[xxx]

## 3.2 EPHMVGA E WINOCCULT

Ephmvga è un programma che mostra le effemeridi per tutti i pianeti più altri corpi celesti. I dati relativi a ciascun corpo celeste comprendono ascensione retta e declinazione relative a ciascuna epoca, l'azimut locale, l'altitudine, le coordinate eliocentriche,tranne per la Luna che sono invece geocentriche, la distanza Terra-Sole, l'elongazione solare, la magnitudine visibile, la percentuale di illuminazione, l'ora locale del sorgere del Sole, del transito al meridiano, e il tramonto, e la distanza angolare tra tutte le combinazioni dei corpi celesti. I tempi sono quello universale e data e ora locale.

Impostando il parametro step size si vede l'andamento nel tempo (scansionato proprio dallo step size) dei dati sopra citati.

Si possono anche visualizzare i movimenti ( in 2D) dei corpi celesti.

Il programma Winoccult è un programma più complesso che viene utilizzato per effettuare previsioni e simulazioni. Ha varie sezioni una riguardante le eclissi e le transiti, un'altra le effemeridi,occultazione di asteroidi.

Effettua previsioni di qualsiasi evento astronomico e lo si può anche visualizzare (in 2D).

## 3.3 PREVISIONE DELL'ECLISSI DI LIBIA

L'eclissi del 29-03-2006 avrà una zona di totalità che sarà visibile da uno stretto corridoio che attraversa mezza Terra (racchiuso in figura 19 dalle linee in viola). La striscia dell'ombra lunare parte dal Brasile, attraversa l'atlantico, il Nord Africa, in Asia centrale e termina a sud-est della

---

[xxx] Sigismondi Costantino, Oliva Pietro, Osservare l'eclissi solari dal bordo della fascia della totalità, articolo dell'Astronomia UAI 2006

Mongolia. L'eclissi parziale sarà vista dalla zona racchiusa tra le linee celesti della figura 18. Questa immagine è stata ottenuta con il programma Winoccult dalla sezione riguardanti le eclissi e inserendo la data (29-03-06).

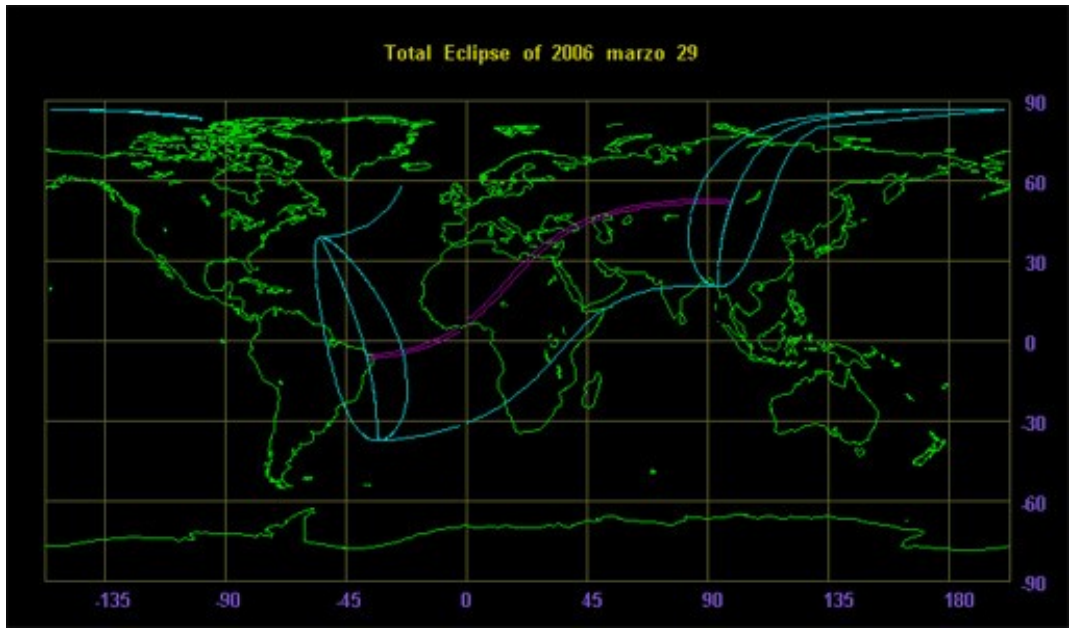

Figura 19 zona di totalità e di parzialità dell'eclisse del 29-03-06: La zona in viola rappresenta la zona terrestre dove si vedrà l'eclissi totale, mentre, nella zona in celeste l'eclisse sarà parziale tanto più ci si allontana dai bordi di totalità.

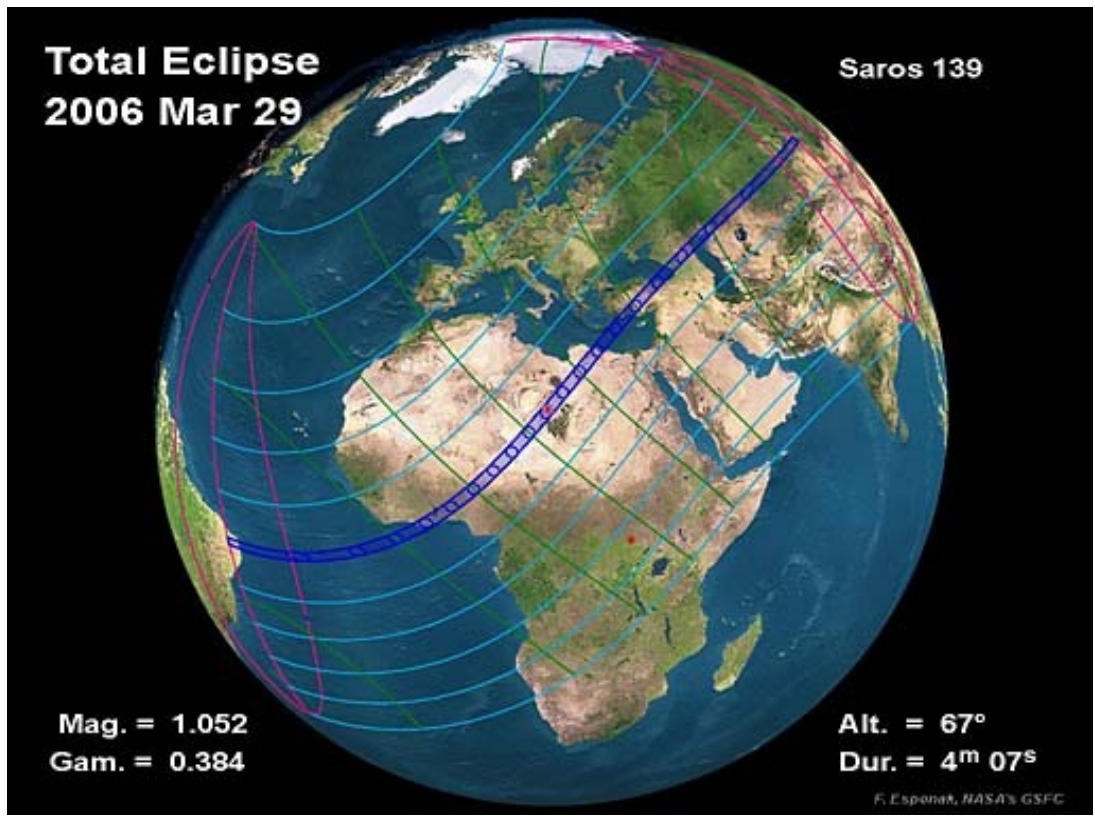

Figura 20 immagine da satellite della zona di totalità e parzialità: La zona in blu scuro è la zona di totalità. Il puntino rosso rappresenta il punto dove l'eclisse durerà più a lungo (4 ' 07'').

Anche in questo caso, come in quello spagnolo, ci sono delle differenze nelle previsioni del bordo di totalitá delle eclissi e della center line. Prendendo i dati di orari e coordinate francesi del bordo di totalità tra la Libia e l'Egitto dell'IMCCE (che utilizzano effemeridi di Chapront per la Luna e Bretagnon per il Sole) facendo il grafico ( figura 21 alla pagina seguente ) si ottiene la previsione francese e le relative equazioni.

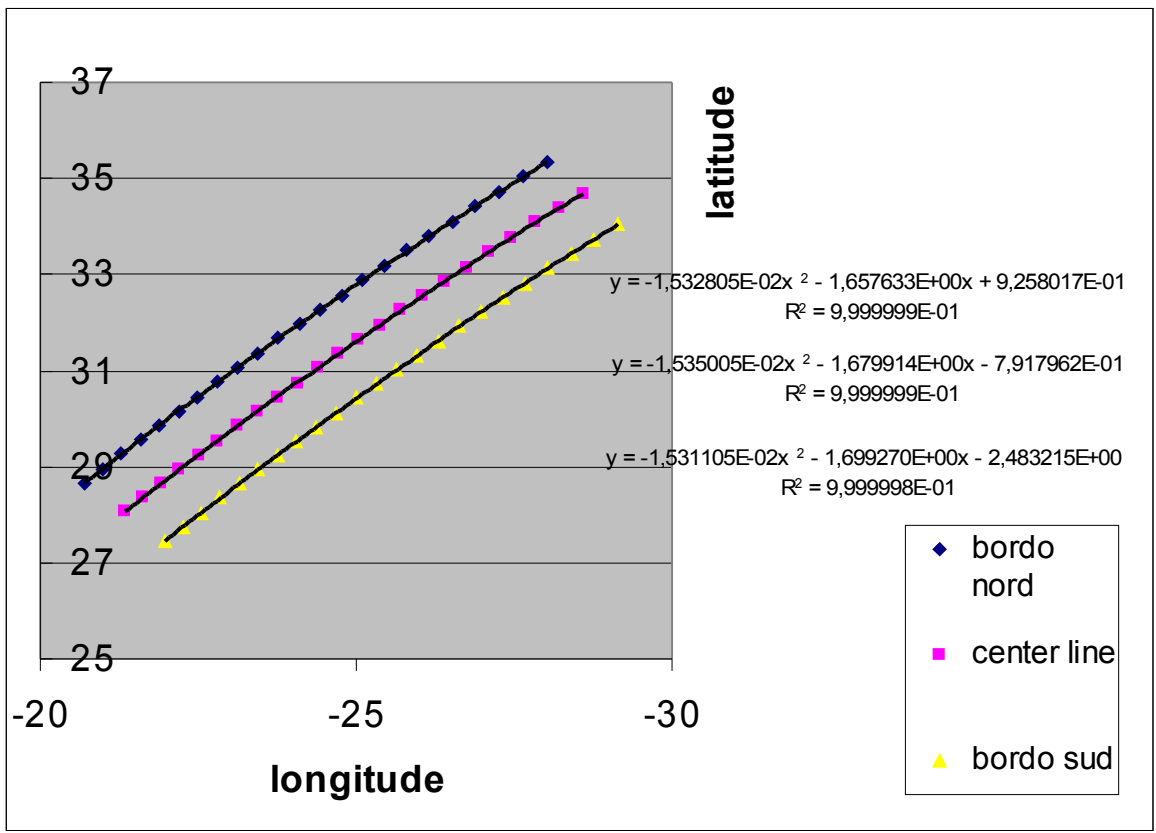

Figura 21 Previsione francese del bordo di totalità: Le tre equazioni in alto a destra riguardano le tre curve (parabole) che si adattano ai dati (curve di regressione). Si è posta la longitudine come variabile indipendente e la latitudine come variabile dipendente per una questione di configurazione.

Successivamente, si sono inserite nelle tre equazioni del bordo francese le coordinate di tre punti corrispondenti al bordo americano (ricavete da un'immagine di Google Maps) e si è ricavato lo spostamento del bordo e della center line rispetto a quello francese.

Vedasi figura 22 alla pagina successiva.

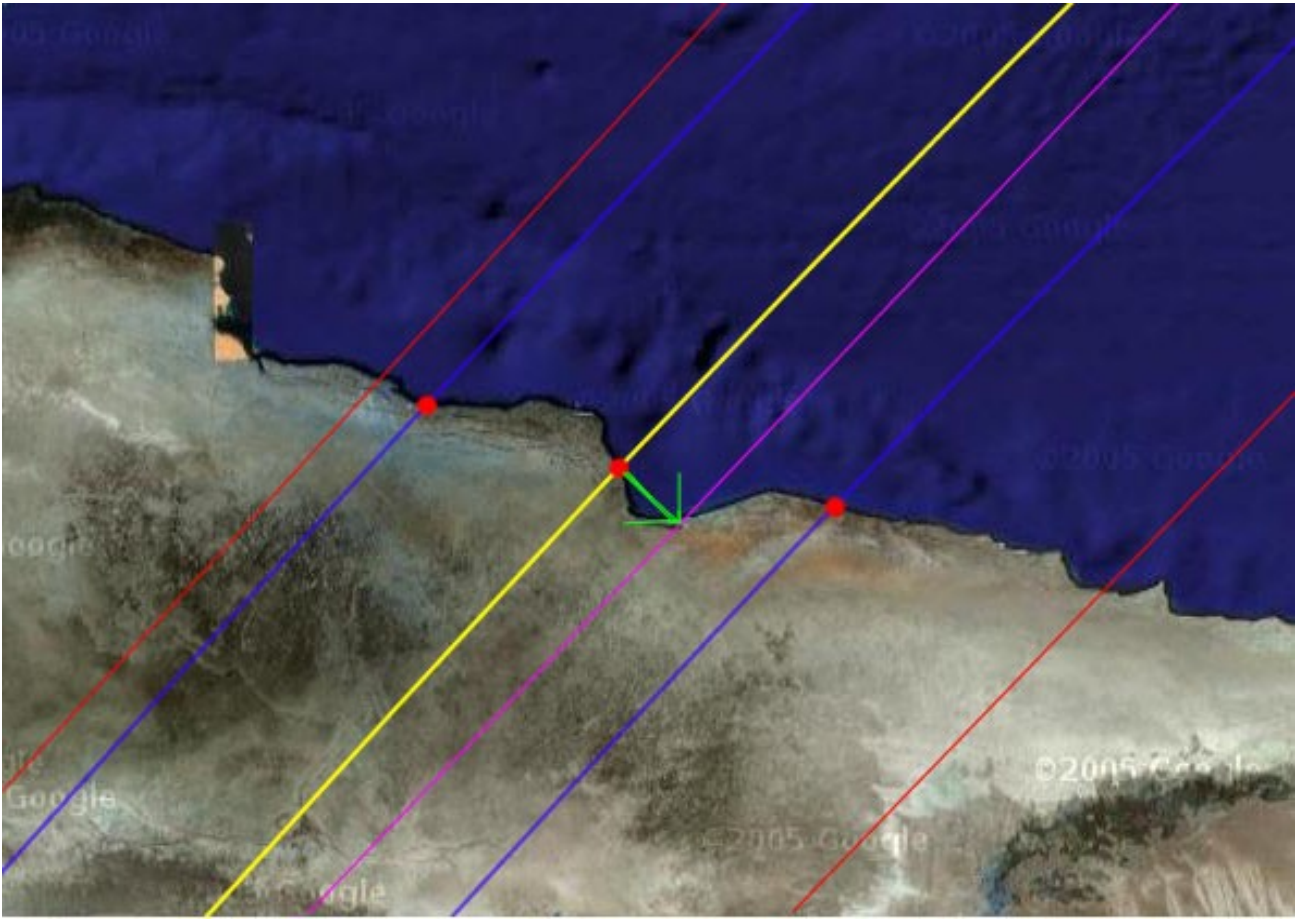

Figura 22 Confronto tra la previsione americana e quella francese: I tre punti rossi sono quelli per cui è stata calcolata la differenza tra le due previsioni. Il bordo blu con la center line gialla è la previsione americana (Espenak, NASA). Il bordo rosso con la center line viola è quello francese (Roche, IMCCE). Si nota dalla figura che il bordo francese è più largo( la center line ed il bordo sud francesi sono più a sud di quelli americani rispettivamente di 1470 e 3560 m mentre il bordo nord é più a nord di 800 m). Questa differenza si spiega nel differente parametro k (raggio medio della Luna espresso in frazione di raggio terrestre).

La NASA utilizza un diverso parametro k (più piccolo $k = 0.2725076$) rispetto a quello consigliato dall'UAI ($k=0{,}272\,5076$) perché prende in considerazione la circonferenza passante per le valli lunari (sicura totalità dell'eclisse a Terra). Ciò che dovrebbe essere uguale nelle due previsioni è la posizione della center line. La correzione effettuata da Roche per tener conto degli scarti tra il centro ottico e quello di massa della Luna è di 0,50'' in latitudine e di -0,25'' in longitudine e per tale correzione si avrebbe uno spostamento complessivo della center line di 986 m. Ma ciò non avviene perché lo spostamento reale è di 1470 m. Avanzano quindi 484 m che derivano da diverse Teorie lunari adottate (effemeridi).

Si può affermare che tale differenza non deriva da diversi sistemi di riferimento utilizzati perché se si localizza la center line francese rispetto a quella americana (facendo variare la latitudine per un dato valore di longitudine), si hanno i seguenti valori:

Tabella 5 Distanza tra le due centerlines:

| |
|---|
| Punto francese 30.7416  Latitudine N |
| Center line -24.6766  Longitudine E |
| Punto americano 30.7577 Latitudine N |
| Center Line -24.46766  Longitudine E |

Riportando tale differenza di latitudine in chilometri si ottiene una distanza di 1,7 Km.

Questa distanza, però, non è misurata sulla perpendicolare tra le due center line, quindi, per ottenerla si moltiplica 1,7 Km per il seno dell'angolo tra questo segmento e la center line americana. Il risultato ottenuto è 1,37 Km a conferma del valore ottenuto in precedenza (1,47 Km).

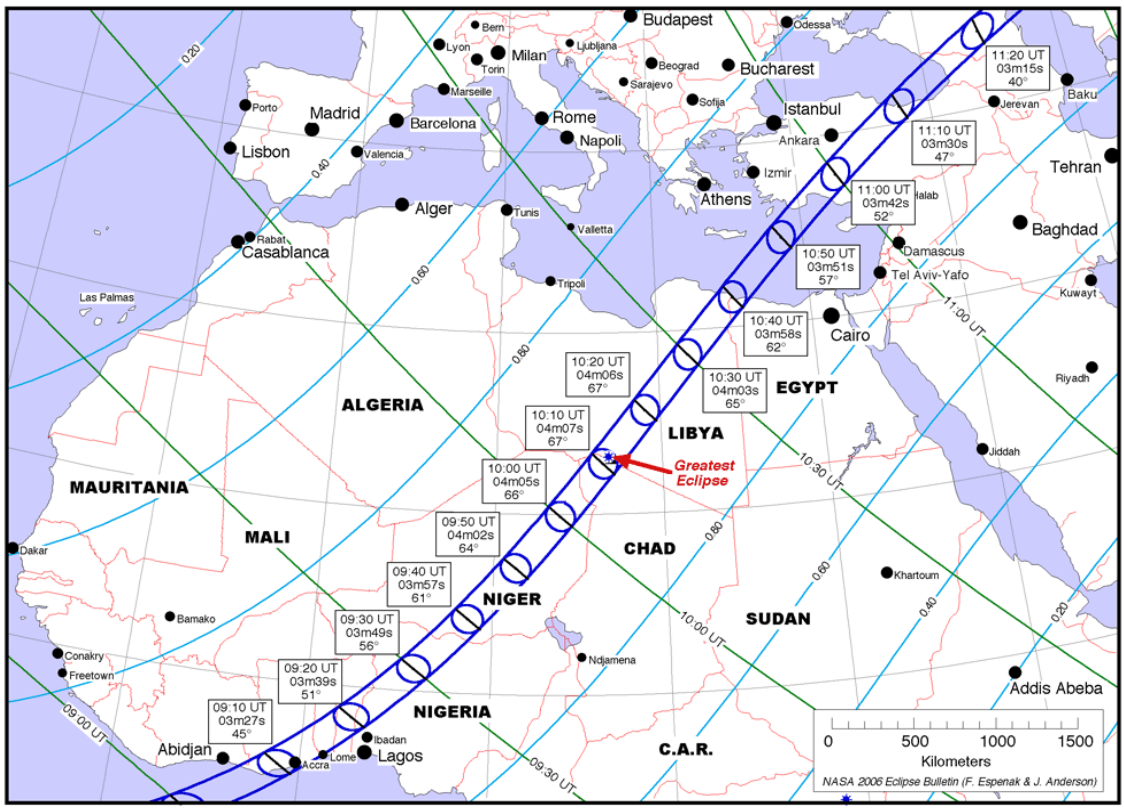

Figura 23. Striscia dell'eclissi totale attraverso l'Africa. Le linee verdi rappresentano l'orario (espresso in Tempo Universale) dell'eclissi in quei punti e nei riquadri è segnata anche la durata. Le linee celesti indicano la percentuale del diametro del Sole (magnitudine) oscurata visibile in quei punti. La fascia blu scuro è quella di totalità.

CONCLUSIONI:

La differenza tra le previsioni NASA e quelle dell'IMCCE di 434±50 m dipendono dal differente uso del parametro k, dalle correzioni della posizione del baricentro lunare e dalle diverse effemeridi usate e non dal diverso sistema di riferimento.

Questo lo si deduce sia dall'esistenza di un sistema di riferimento globale, cioè il WGS 84, e sia dal fatto che la discrepanza in latitudine derivante da diversi sistemi di riferimento si aggira all'incirca da 1" ai 6" che equivalgono a 30 m e a 180 m; valori comunque abbastanza inferiori rispetto ai quasi 500 m calcolati.

Con il LLR (Lunar Laser Ranging) e con misure accurate effettuate durante l'eclissi (sul bordo di totalità) si possono migliorare i risultati ottenuti finora. Si possono ridurre le incertezze sulle correzioni al baricentro che simulazioni di WinOccult sono ancora parametri liberi.

## Indice delle tabelle e delle figure

# Bibliografia e siti Web